\documentclass{aa} 
\usepackage{graphicx}
\usepackage{lscape}
\usepackage{subfigure}
\usepackage{natbib}
\usepackage{hyperref}
\usepackage[varg]{txfonts}
\usepackage{longtable}
\usepackage{booktabs}
\usepackage{enumerate}

\bibpunct{(}{)}{;}{a}{}{,} 
\begin{document}

   \title{C {\Large{IV}} BAL disappearance in a large SDSS QSO sample}

   \author{D. De Cicco \inst{1}, W.~N. Brandt\inst{2,3,4}, C.~J. Grier\inst{2}, M. Paolillo\inst{1,5,6}, N. Filiz Ak\inst{7,8}, D. P. Schneider\inst{2,3}, J. R. Trump\inst{9}}
   \titlerunning{BAL variability in quasar spectra}
   \authorrunning{D. De Cicco et al.}

 \institute{Department of Physics, University of Napoli ``Federico II'', via Cinthia 9, 80126 Napoli, Italy\\e-mail: demetradecicco@gmail.com, demetra.decicco@unina.it 
 	\and
 		Department of Astronomy and Astrophysics, 525 Davey Laboratory, The Pennsylvania State University, University Park, PA 16802, USA 
	\and
		Institute for Gravitation and the Cosmos, The Pennsylvania State University, University Park, PA 16802, USA
	\and
		Department of Physics, 104 Davey Laboratory, The Pennsylvania State University, University Park, PA 16802, USA 
        \and
      		INFN - Sezione di Napoli, via Cinthia 9, 80126 Napoli, Italy 
	\and
		ASI Science Data Center, via del Politecnico snc, 00133 Roma, Italy 
	\and
		Faculty of Sciences, Department of Astronomy and Space Sciences, Erciyes University, 38039, Kayseri, Turkey 
	\and
		Astronomy and Space Sciences Observatory and Research Center, Erciyes University, 38039, Kayseri, Turkey 
	\and
		University of Connecticut, Storrs, CT 06269, USA 
	}

   \date{}
  \abstract
   {Broad absorption lines (BALs) in the spectra of quasi-stellar objects (QSOs) originate from outflowing winds along our line of sight; winds are thought to originate from the inner regions of the QSO accretion disk, close to the central supermassive black hole (SMBH). These winds likely play a role in galaxy evolution and are responsible for aiding the accretion mechanism onto the SMBH. Several works have shown that BAL equivalent widths can change on typical timescales from months to years; such variability is generally attributed to changes in the covering factor (due to rotation and/or changes in the wind structure) and/or in the ionization level of the gas.}
   {We investigate BAL variability, focusing on BAL disappearance.}
   {We analyze multi-epoch spectra of more than 1500 QSOs --the largest sample ever used for such a study-- observed by different programs from the Sloan Digital Sky Survey-I/II/III (SDSS-I/II/III), and search for disappearing C {\scriptsize{IV}} BALs. The spectra cover a rest-frame time baseline ranging from 0.28 to 4.9 yr; the source redshifts range from 1.68 to 4.27.}
   {We detect 73 disappearing BALs in the spectra of 67 unique sources. This corresponds to 3.9\% of BALs disappearing within 4.9 yr (rest frame), and 5.1\% of our BAL QSOs exhibit at least one disappearing BAL within 4.9 yr (rest frame). We estimate the average lifetime of a BAL along our line of sight ($\approx 80-100$ yr), which appears consistent with the accretion disk orbital time at distances where winds are thought to originate. We inspect various properties of the disappearing BAL sample and compare them to the corresponding properties of our main sample. We also investigate the existence of a correlation in the variability of multiple troughs in the same spectrum, and find it persistent at large velocity offsets between BAL pairs, suggesting that a mechanism extending on a global scale is necessary to explain the phenomenon. We select a more reliable sample of disappearing BALs on the basis of the criteria adopted by \citet{nur12}, where a subset of our current sample was analyzed, and compare the findings from the two works, obtaining generally consistent results.}
   {}


   \maketitle
   
%

\section{Introduction}
\label{section:intro}
QSO spectra are characterized by prominent emission lines originating from ultraviolet (UV) transitions, such as N {\footnotesize{V}}, C {\footnotesize{IV}}, Si {\footnotesize{IV}}, down to lower ionization lines, such as Al {\footnotesize{III}}, Mg {\footnotesize{II}} \citep[e.g.,][]{vandenBerk}. In $10\%-20\%$ of optically selected QSOs, absorption features corresponding to the aforementioned higher-ionization transitions are also observed \citep[e.g.,][]{hf,g09,allen}; they are typically blueshifted up to 0.2\emph{c} from the corresponding rest-frame line (but rare examples of QSO spectra exhibiting redshifted absorption lines are also known; see, e.g., \citealt{hall13}). In $\approx 10\%$ of the cases, additional features corresponding to the lower ionization transitions are observed \citep[e.g.,][]{weymann91, murray}.

The presence of absorption lines suggests that some mechanism exists allowing the transfer of a significant amount of momentum from the radiation field to the gas where the lines originate. Absorption features are thought to arise from radiatively accelerated winds which, in turn, originate from the inner region of the accretion disk surrounding the central supermassive black hole (SMBH; typical distances are on the order of $10^{-2}-10^{-1}$ pc; e.g., \citealt{murray,proga}).

Winds likely enable the accretion mechanism by removing from the disk the angular momentum carried by the accreting material. Moreover, they may affect star formation processes and hence galaxy evolution as a whole since they evacuate gas from the host galaxy and redistribute it in the intergalactic medium \citep[e.g.,][]{dimatteo, capellupo}; in addition, they can prevent new gas inflow into the galaxy (a process known as ``strangulation''; see, e.g., \citealt{peng}, and references therein).

Several models proposed to describe QSO winds \citep[e.g.,][]{elvis,proga} suggest that the observation of absorption lines depends on the viewing angle, thus providing a possible explanation for the lack of detection of absorption lines in most QSO spectra; alternatively, we could envision absorption lines as the signature of a specific phase in QSO evolution \citep[e.g.,][]{green}. 

The detection of spectral features originating in the proximity of the SMBH is somewhat surprising: QSOs are typically extremely luminous, having a bolometric luminosity on the order of $10^{46}$ erg s$^{-1}$, and most of the energy they radiate consists of strong UV and X-ray emission from the inner regions; such emission is therefore expected to overionize the gas it interacts with at small distances and, as a consequence, no spectral lines should be observed at all \citep[e.g.,][]{proga}. Thus, any model attempting to describe the origin and effects of QSO winds must provide a solution for the ``overionization problem'' and explain why the gas is not fully ionized. The presence of shielding material located between the central source of radiation and the outflowing winds could possibly account for the observed lines \citep[e.g.,][]{murray}. \citet{baskin} propose an alternative model, based on the results of hydrostatic simulations. The ionization state of winds can be defined by means of the ionization parameter $U = n_\gamma/n_e$, where $n_\gamma$ is the density of ionizing photons and $n_e$ is the electron density. Essentially, the model assumes highly clumped winds, due to the compression exerted by the ionizing radiation; this situation leads to high electron densities and thus to an ionization parameter that is low enough to prevent overionization, but sufficiently ionized to allow the formation of the observed UV spectral lines. According to the model, the outflowing gas is compressed in the radial direction, and thus forms sheets or filaments along the line of sight; this gives rise to density gradients, and hence to different ionization states of the outflowing gas along the line of sight.

It is common practice to characterize absorption lines in terms of velocity: the two extremes of a line define the maximum blueshift velocity $v_{\mbox{\tiny{max}}}$ and the minimum blueshift velocity $v_{\mbox{\tiny{min}}}$; this allows definition of the central velocity $v_{\mbox{\tiny{c}}} = (v_{\mbox{\tiny{max}}} + v_{\mbox{\tiny{min}}})/2$, which is an indicator of the position of the feature, and the velocity difference $\Delta v = |v_{\mbox{\tiny{max}}} - v_{\mbox{\tiny{min}}}|$, which defines the width of the absorption line.
Traditionally, absorption lines are classified on the basis of their width $\Delta v$ in velocity units: broad absorption lines (BALs) if $\Delta v \geq 2000$ km s$^{-1}$, mini-BALs when $500 \leq \Delta v < 2000$ km s$^{-1}$, and narrow absorption lines (NALs) for $\Delta v < 500$ km s$^{-1}$. In turn, BAL QSOs are classified into three groups depending on the observed transitions: HiBALs exhibit only absorption features from highly-ionized species, such as C {\footnotesize{IV}}, Si {\footnotesize{IV}}, N {\footnotesize{V}}; LoBALs are characterized by lower-ionization lines, such as Al {\footnotesize{II}}, Al {\footnotesize{III}}, Mg {\footnotesize{II}}, in their spectra, in addition to the above-mentioned high-ionization absorption features. When iron lines, such as Fe {\footnotesize{II}} and Fe {\footnotesize{III}}, are observed in addition to both high- and low-ionization lines, QSOs are referred to as FeLoBALs \citep[e.g.,][]{hazard,vwk}. 

Over the last decades BALs have been extensively studied with the aim of gaining insights into the geometry and physics of QSOs, their associated winds, and the emission/absorption processes that characterize them. BAL troughs are detected in $\approx15\%$ of QSO spectra and by definition lie at least $10\%$ below the continuum \citep[e.g.,][and references therein]{weymann91, trump, g09}.

Since the 1980s, several studies have revealed that the equivalent widths (EWs) of BAL troughs can vary on typical rest-frame timescales ranging from months to years (but occasionally much shorter; see, e.g., \citealt{grier15}, where the variability of a C {\footnotesize{IV}} BAL trough on rest-frame timescales of $\approx 1.20$ days is discussed).

According to the most successful theories, BAL variability originates from changes in the features (such as density, geometry) of the absorber, which --in turn-- are due to rotation and/or variations in the characteristics of the outflowing winds \citep[e.g.,][]{proga}; this model leads to variations in the covering factor (i.e., the fraction of radiation emitted by the central engine that is blocked by the absorbing material) along the line of sight, depending on the velocity structure of the outflows. Variations of the absorber could also involve changes in the ionization level of the gas which could manifest themselves as absorption-line variations; the ion abundances vary, causing the weakening/strengthening of the absorption lines \citep[e.g.,][]{B93}. The investigation of BAL variability thus reveals information about the structure and dynamics of the outflowing winds, hence constraining QSO outflows. 

Previous studies of BAL variability were generally limited by the difficulty in obtaining multi-year observations for large samples of sources; typically either the sample size or the observing-baseline length were sacrificed. In order to report some remarkable examples, \citet{B93} monitored a sample of 23 QSOs by means of observations covering a $\approx 1$ yr timescale, and variability was detected in 15 sources; \citet{L07} investigated C {\footnotesize{IV}} BAL variability over a $< 1$ yr baseline in a QSO sample consisting of 29 objects, while \citet{g08} characterized C {\footnotesize{IV}} BAL variability in 13 QSOs observed over $3-6$ yr.\footnote{All the mentioned timescales are in the rest frame.} Some works report BAL disappearance \citep[e.g.,][]{L07} or emergence \citep[e.g.,][]{capellupo} instances in individual sources. 

\citet{nur12} presented the first statistical analysis of C {\footnotesize{IV}} BAL disappearance. Their study draws from an ancillary project making use of data from different surveys that are part of the Sloan Digital Sky Survey-I/II/III (SDSS-I/II/III; e.g., \citealt{sdss}), and is based on a sample consisting of 582 QSOs with rest-frame observing baselines ranging from $1.1-3.9$ yr and with spectral epochs from at least two surveys; this allows the investigation of BAL variability by means of the comparison of at least two spectra of the same source. 

The analysis we shall present takes its cue from \citet{nur12} and takes advantage of observations from the Baryon Oscillation Spectroscopic Survey (BOSS; e.g., \citealt{boss}) that were not available at the time that work was published, thus significantly enlarging (more than double) the sample of inspected sources. This is the first time that such a large sample has been used for BAL disappearance analysis.

The paper is organized as follows: in Section \ref{section:dataset} we introduce in detail our sample of BAL QSOs; in Section \ref{section:reduction} we describe the spectral reduction process and discuss the method used to investigate C {\footnotesize{IV}} BAL disappearance; in Section \ref{section:analysis} we present the results of our analysis, and we discuss our findings and draw some conclusions in Section \ref{section:discussion}. Throughout the present work, timescales and lifetimes are in the rest frame, and when discussing disappearing BAL troughs we will always refer to C {\footnotesize{IV}}, unless otherwise stated.

\section{Observations and sample selection}
\label{section:dataset}
\subsection{The Baryon Oscillation Spectroscopic Survey}
\label{section:boss}
Our analysis of C {\footnotesize{IV}} BAL disappearance is based on data from BOSS, which is the largest of the four SDSS-III surveys \citep{eisenstein}. BOSS was designed to map the baryon acoustic oscillation (BAO) signature imprinted in QSOs and luminous red galaxies by tracing their spatial distribution, aiming to measure cosmic distances with the ultimate goal of providing improved constraints on the acceleration of the expansion rate of the Universe. 

Observations were made with the 2.5 m SDSS telescope \citep{gunn} at Apache Point Observatory, New Mexico, USA. The telescope is equipped with two identical spectrographs, each with one camera for red and one for blue wavelengths, covering the wavelength ranges $5650-10000$ \AA$\mbox{ }$and $3600-6350$ \AA, respectively \citep{smee}. The spectral resolution varies within the range $1850-2650$ in the red channel and $1560-2270$ in the blue channel. Spectra of numerous sources are acquired simultaneously by means of aluminium plates subtending a 3 degree-wide area on the sky and attached to the telescope; 1000 holes are drilled on each plate, in correspondence with the position on the sky of the sources to be observed, and a 2\arcsec-diameter fiber is plugged into each hole. Light from each observed source is directed to a dichroic that splits the red and blue components of the spectrum, so that each of them is registered on a different CCD. 

BOSS observations spanned four and a half years (Fall 2009 $-$ Spring 2014) and surveyed more than $10000$ square degrees previously investigated by the SDSS-I/II surveys; approximately 1.5 million luminous galaxies with a redshift in the range $0.15 - 0.7$, together with $\approx 184000$ QSOs with $2.15 \leq z \leq 3.5$, were targeted \citep{paris}. 

SDSS-I/II spectra cover the wavelength range $3800-9200$ \AA, while BOSS spectra extend further in each direction to cover the range $3650-10400$ \AA.

\subsection{Selection of QSOs and spectra}
\label{section:bal_sample}
SDSS-I/II allowed the identification of thousands of BAL QSOs. \citet{g09} presented a catalog of 5039 of them, selected from the SDSS Data Release 5 QSO catalog \citep{schn07}. The BOSS survey retargeted, among other sources, 1606 of these QSOs drawn from a sample of 2005, to allow for investigation of BAL variability on multi-year timescales, in order to gain knowledge of the structure, dynamics and physical properties of QSO winds \citep{boss, nur13}. The selected QSOs fulfill the following requirements:
\begin{itemize}
\item[--] are optically bright, having an $i$ band magnitude $< 19.3$ mag;
\item[--] show strong BALs, i.e., have a balnicity index (following the definition by \citealt{g09}; see below) BI$_0 > 100$ km s$^{-1}$ in at least one of the observed BAL troughs; 
\item[--] spectra are characterized by a high signal to noise ratio (S/N; specifically, in \citealt{g09} it is denoted as SN$_{1700}$, and is required to be $> 6$; see below).
\end{itemize}
The balnicity index was proposed by \citet{weymann91} in order to quantify the BAL nature of a QSO by means of a continuous indicator characterizing the C {\footnotesize{IV}} absorption lines in a spectrum; it was defined as 
\begin{equation}
BI = -\int_{25000}^{3000}[1-f(v)/0.9]C dv\mbox{  ,}
\label{eqn:BI}
\end{equation}
where $f(v)$ is the continuum-normalized flux as a function of the velocity displacement with respect to the line center, while $C$ is a constant that is set to 0 and switches to 1 only when the quantity in the brackets is continuously positive over a velocity range that is $\geq 2000$ km s$^{-1}$. Such a definition allows measurement of the EW of an absorption line (in km s$^{-1}$) provided the line is broader than 2000 km s$^{-1}$ and is at least 10\% below the continuum. The 25000 km s$^{-1}$ blue limit and the 3000 km s$^{-1}$ red limit were set to avoid contamination from Si {\footnotesize{IV}} emission/absorption features on the blue end of the line and from the C {\footnotesize{IV}} emission line on the red end. \citet{g09} further defined a modified balnicity index BI$_0$, where the red-end limit is replaced by 0 km s$^{-1}$.

SN$_{1700}$ was introduced by \citet{g09} to quantify the S/N corresponding to the measurement of C {\footnotesize{IV}} absorption; it is defined as the median of the flux divided by the noise (provided by the SDSS pipeline), and is computed for the spectral bins in the wavelength range $1650-1750$ \AA, which was chosen because it is relatively close to the C {\footnotesize{IV}} BAL region and usually characterized by little absorption. The binning is the one provided in the SDSS spectra. Several factors (e.g., the integration time or the source redshift and luminosity) can affect the estimate of SN$_{1700}$; nevertheless, it allows selection of QSO samples on the basis of their S/N.

Our analysis focuses on C {\footnotesize{IV}} BALs since these are the most commonly observed \citep[e.g.,][and references therein]{g09, wildy}; in addition, contamination by adjacent absorption features is generally not significant.

The disappearance of a BAL can be detected by comparing two or more spectra of the same QSO taken at different times; as a consequence, we require sources with spectral coverage both from BOSS and from previous SDSS-I/II surveys. This requirement restricts our sample to 1606 out of 2005 QSOs, since there are no BOSS spectra for the rest of the sources. We obtain BOSS spectra from the SDSS Data Release 12, selecting a matching radius of 0.0005 deg with the coordinates from our source list. A further restriction is necessary in order to select the redshift window of interest: $1.68 < z < 4.93$ for C {\footnotesize{IV}} BALs to be fully visible in SDSS spectra, considering that their blueshifted velocities can range from $-30000$ to 0 km s$^{-1}$ \citep{g09}; this requirement cuts our sample of QSOs down to 1525 sources. We make use of redshifts from \citet{hw}.

We convert wavelengths to velocities and, following other works from the literature \citep[e.g.,][]{nur12}, we require our BALs to have a maximum velocity $-30000 \leq v_{\mbox{\tiny{max}}} \leq -3000$ km s$^{-1}$ --and thus exclude all the BALs entirely confined within the range $-3000 - 0$ km s$^{-1}$-- on the basis of what stated above about contamination by other features. \citet{nur12} showed that setting $v_{\mbox{\tiny{max}}} \leq -3000$ km s$^{-1}$ does not introduce any significant bias in their results, as disappearing BALs are generally characterized by high absolute values of $v_{\mbox{\tiny{min}}}$, and will therefore have even higher absolute values of $v_{\mbox{\tiny{max}}}$; this behavior is also apparent in the present work when analyzing the $v_{\mbox{\tiny{min}}}$ distribution of the BALs in our sample (see Section \ref{section:vel}).

Our final sample, selected adopting a conservative approach, consists of all the QSOs whose SDSS-I/II spectra show at least one C {\footnotesize{IV}} BAL in the velocity range of interest, that is to say, 1319 sources (hereafter, main sample); thus, 206 sources from the 1525 we selected on the basis of their redshift are dropped from the sample. Since the 206 objects, as well as the others, belong to the sample of BAL QSOs of \citet{g09}, we inspect the available spectra for each source, and compare them to the spectra obtained by \citet{g09}, in order to understand why we do not find BALs for them. Here we list the causes for such a discrepancy:

\begin{itemize}
\item[--] in 123 instances the BALs in \citet{g09} are low-velocity absorption lines, close to the C {\footnotesize{IV}} emission line; in our spectra we detect these absorption lines, but the fraction below 10\% of the continuum level is much narrower than the required 2000 km s$^{-1}$; of course this result is understandable, given the vicinity of the emission line contaminating the absorption; \citet{g09} point out that, in such cases, the characterization of emission lines is particularly problematic, and that low-velocity BAL trough measurements are therefore less certain than the others. Adhering to our conservative approach, we exclude these sources from our analysis;
\item[--] in 25 instances we do detect BALs as in \citet{g09}, but they have $v_{\mbox{\tiny{max}}} > -3000$ km s$^{-1}$, so we reject them because of the aforementioned requirement on the maximum velocity;
\item[--] in 22 instances we do not detect BALs in our spectra at locations given in \citet{g09}: this change is due to small differences in the normalized continua, and to BALs being very shallow;
\item [--] in 17 instances the BALs detected by \citet{g09} are very blueshifted and are out of the velocity range of interest, having $v_{\mbox{\tiny{max}}} < -30000$ km s$^{-1}$;
\item[--] in ten instances, when we try to compare the SDSS-I/II and BOSS spectra of the same source, we find that they are characterized by continua that do not overlap due to a vertical shift, hence comparing them would be improper; this generally depends on the continuum fit of the blue end of the spectra. In principle we could shift one of the two spectra to obtain overlapping continua; nevertheless, consistently with our conservative approach, we prefer to exclude these sources from the sample we analyze. We will mention again these ten sources in Section \ref{section:discussion}, where we will briefly discuss the effect of a potential inclusion in our sample;
\item[--] in eight instances we do not detect BALs because of noise spikes that split them into two or more mini-BALs, so technically they do not fulfill the BAL defining criteria (these sources include three classified as BAL QSOs in \citealt{nur12}, as detailed in Section \ref{section:nur}; their spectra will be shown in Fig. \ref{fig:fa_bals});
\item[--] in one instance we are unable to fit the spectra properly because of a flux drop at the blue end; moreover, the estimate of the source redshift is uncertain.
\end{itemize}

Our sample of 1319 sources is a factor of 2.3 times larger than the sample inspected in \citet{nur12}. For each QSO, at least a pair of spectra --one from SDSS-I/II and one from BOSS observations-- is available; 343 sources have one or more additional spectra from SDSS-I/II and/or BOSS, yielding three or more total epochs for these objects. Throughout the present work, unless otherwise stated, we will label as a ``pair of spectra'' (or, equivalently, a ``pair of epochs'') two spectra of the same source, where the earlier spectrum is from SDSS-I/II and the later one is from BOSS. The rest-frame timescales between observations in a pair are in the range $0.28-4.9$ yr. 

\section{Spectral data processing}
\label{section:reduction}
BOSS spectra can be affected by systematics due to spectrophotometric calibration errors \citep{margala}. These uncertainties mostly arise from the differential refraction of light in the atmosphere and from a focal-plane offset of the positions of the fibers targeting QSOs with respect to the ones assigned to calibration stars; as a result, we obtain a higher throughput in the Lyman-$\alpha$ wavelength window when we observe high-redshift QSOs. This leads to a higher S/N at a cost of larger calibration errors with respect to those in SDSS-I/II spectra. We make use of the corrections implemented and discussed by \citet{margala} to remove such errors.

 The header files of SDSS spectra include bitmasks quantifying the ``goodness'' of each pixel constituting the detector, based on a set of observational and instrumental conditions. SDSS spectra are generally obtained by the combination of three or more exposures, and there may be pixels whose goodness changes from one exposure to another; the \emph{and\_mask} column in the Header Data Unit identifies the pixels that are bad in all exposures. Following \citet{nur12} and \citet{grier16}, we mask all the pixels that are flagged as bad with respect to the ``BRIGHTSKY'' threshold, as this indicates that the flux contribution from the sky in such pixels is too high.

\subsection{Extinction correction}
\label{section:ext_corr}
Spectra must be corrected for Galactic extinction before use. \citet{cardelli} derive an extinction law $A(\lambda)/A(V)$, where $A(\lambda)$ is the absolute extinction at the wavelength of interest and $A(V)$ is the absolute visual extinction; historically, the \emph{V} band is used as a reference. This extinction law is valid in the wavelength range $0.125 \leq \lambda \leq 3.5 \mbox{ }\mu$m, and depends on the parameter $R_{\mbox{v}}= A_V/E(B-V)$, i.e., the ratio of visual extinction to reddening. We follow \citet{cardelli} to correct for Galactic extinction, and adopt a Milky Way extinction model, where we assume $R_{\mbox{v}}=3.1$; visual extinction coefficients are from \citet{schlegel}. Once the correction is performed, we convert the observed wavelengths to rest-frame wavelengths and to velocities.

\subsection{Continuum fit}
\label{section:cf}
In the process of fitting a continuum model to our spectra we follow the steps described by \citet{grier16} and \citet{g09}. We adopt a reddened power law model and make use of the Small Magellanic Cloud-like reddening curve presented and discussed by \citet{pei}. 

We identify a set of relatively line-free (RLF) regions --i.e., regions which are generally free from strong emission/absorption features-- to fit the continuum of each spectrum. The same set of RLF regions is used for each spectrum, although in some cases one or more regions may be not available in a spectrum. The selected wavelength windows are $1280-1350$, $1425-1450$, $1700-1800$, $1950-2200$, $2650-2710$, $3010-3700$, $3950-4050$, $4140-4270$, $4400-4770$, and $5100-6400\mbox{ }\AA$. Most of them have been used in several works from the literature \citep[e.g.,][]{g09,nur12}; here, following \citet{grier16}, we reduced the usual width (i.e., $1250-1350\mbox{ }\AA$) of the first region in the list on its blue side, in order to limit contamination from possible nearby emission features; moreover, we introduced an additional RLF window corresponding to the wavelength range $1425-1450$ \AA, to be used if the source redshift is $z < 1.85$, in order to obtain a better fitting of the blue end of the corresponding spectrum.

We assign each of the regions used equal weight in the fitting process regardless of the number of pixels from which it is composed, hence we attribute a weight to each pixel in each RLF region on the basis of the total number of pixels constituting the region itself. 

The continuum fitting is always a challenging task, as no wavelength region is completely free of emission/absorption lines. In order to minimize the risk of contamination by prominent features happening to fall in the selected RLF regions, we fit our continuum through a non-linear least square analysis performed iteratively, and set a $3\sigma$ threshold to exclude outliers at each iteration. We then visually inspected each RLF region in each spectrum after the iterative fitting, and made sure that troubled regions had no significant effect in the fitting process. Following \citet{baskin13}, we tested the reliability of our results performing an alternative continuum fit using only one RLF region blueward of C {\footnotesize{IV}}, corresponding to $1275-1285\mbox{ }\AA$, which is expected to be unaffected by emission/absorption features. The results of the corresponding analysis of BAL disappearance are consistent with the ones we are about to describe, confirming that our continuum fitting procedure is robust with respect to the presence of possible contaminating features.

We quantify uncertainties in the continuum fit by means of Monte Carlo simulations performed iteratively, following \citet{peterson} and \citet{grier16}: essentially, we alter the flux of each pixel in the spectrum by a random Gaussian deviate multiplied by its uncertainties; 
we add random noise based upon each pixel's error estimate with a Gaussian distribution to each spectrum, fit the continuum to the obtained spectrum, and iterate the procedure 100 times. The standard deviation of the 100 iterations is assumed to be the uncertainty of the continuum fit for that spectrum.

\begin{figure*}[h!]
 \center
   {\includegraphics[width=15cm]{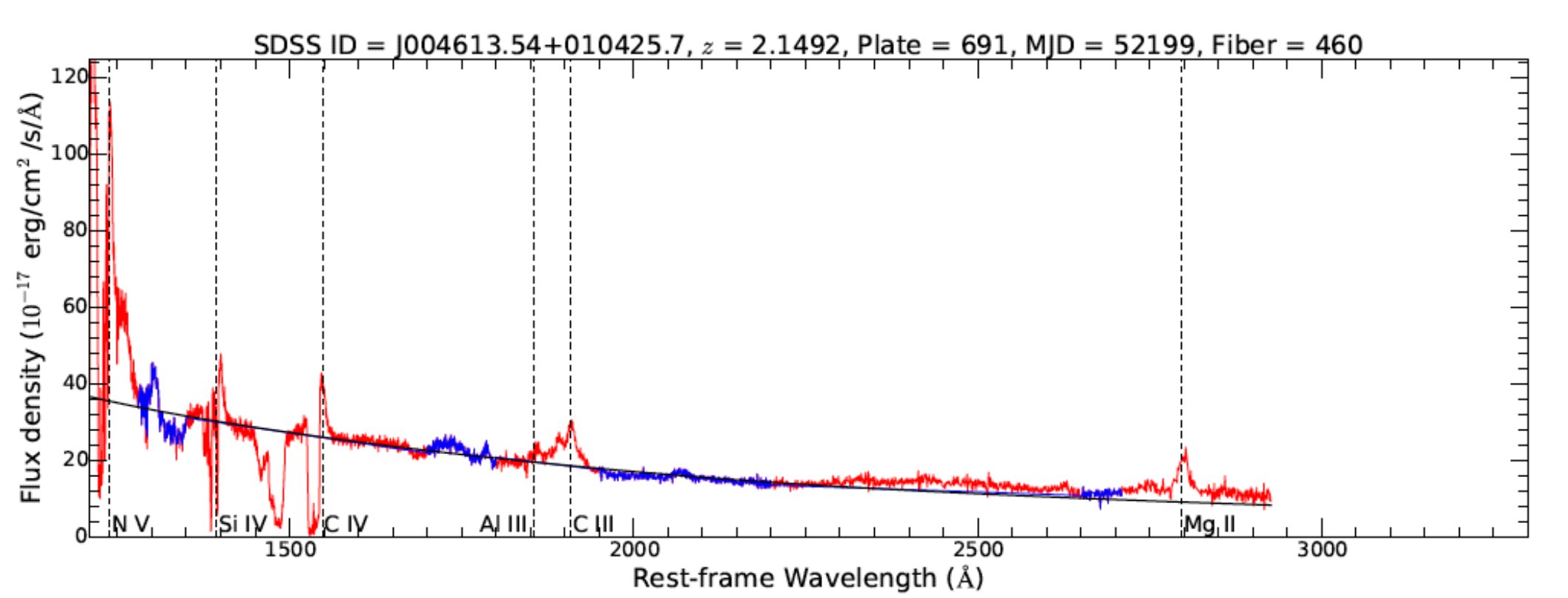}}
   {\includegraphics[width=15cm]{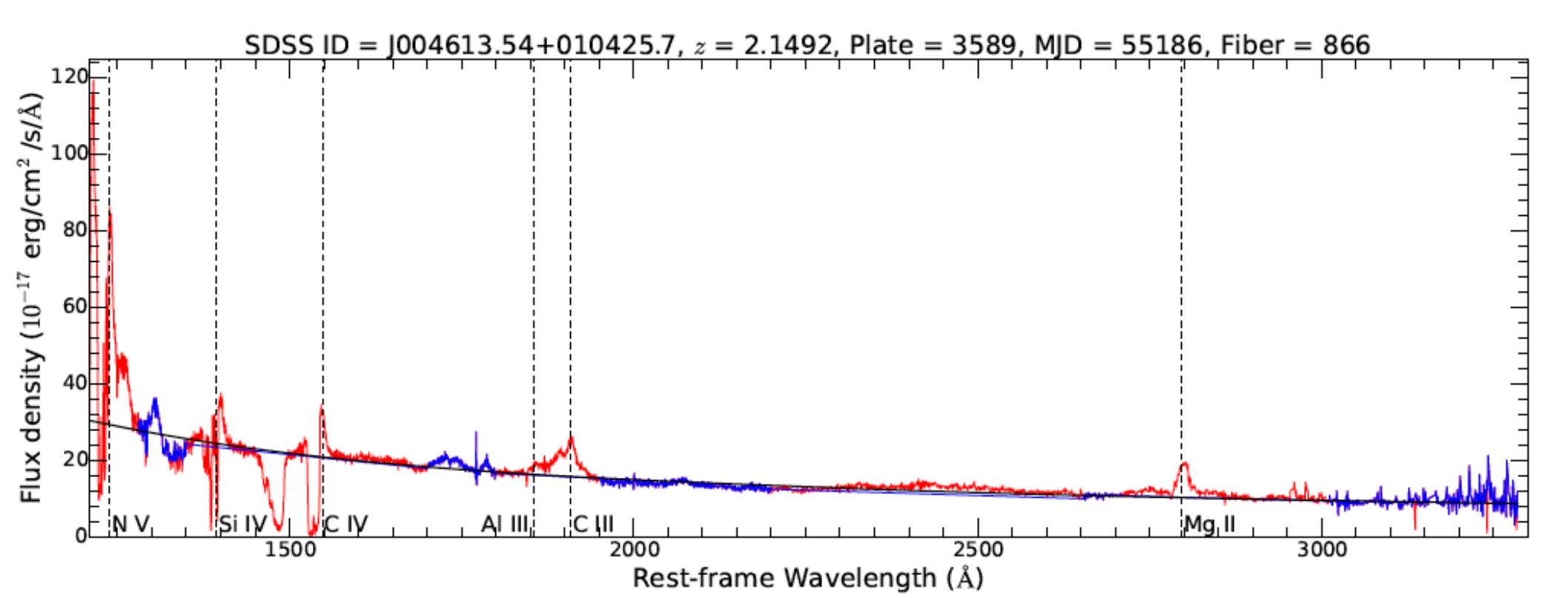}}
\caption[Examples of continuum fits]{\footnotesize Examples of continuum-model fits for a pair of spectra (SDSS-I/II in the upper panel, BOSS in the lower panel) where we observe C {\scriptsize{IV}} BAL disappearance. The red line represents the signal as a function of the rest-frame wavelength, while the blue regions indicate the RLF windows used to fit the continuum model, which is shown by the solid black line. Vertical dashed lines indicate rest-frame wavelengths corresponding to features that are typically seen in QSOs. The source ID in the SDSS archive, together with its redshift and information about the plate, MJD, and fiber corresponding to the observation, are also reported on top of each panel.}\label{fig:spectra}
\end{figure*}

As mentioned above (see Section \ref{section:boss}) SDSS-I/II and BOSS spectra have different spectral coverage, so we crop BOSS spectra to match the SDSS-I/II wavelength coverage, thus ensuring that the same RLF regions are used in the continuum fit process for all the spectra corresponding to each source. Figure \ref{fig:spectra} displays a pair of spectra where disappearing C{~}{\footnotesize{IV}} BAL troughs are observed; RLF regions and the best-fit continuum model for each spectrum are also shown, together with the main emission features that are typical of BAL QSOs in the observed wavelength range. 

SDSS spectra are identified by three integer numbers: plate, modified Julian date (MJD), and fiber; they identify the aluminium plate used to obtain the spectrum, the observing night, and the fiber used to observe the source of interest, respectively. Fiber number ranges from $1-1000$ for BOSS observations, and $1-640$ for SDSS-I/II observations, as a smaller number of fibers was used at the time (see \citealt{smee}). Plate, MJD, and fiber numbers are reported on top of each panel in Fig. \ref{fig:spectra}. 

\section{Statistical analysis of the disappearing BAL sample}
\label{section:analysis}
\subsection{Identification of disappearing BAL troughs}
\label{section:disapp}
The goal of the present work is to investigate the disappearance of C {\footnotesize{IV}} BAL troughs in the largest available sample of BAL QSOs as well as the existence of coordination in the variability of multiple troughs corresponding to the same transition in a spectrum. The large size of our sample allows us to perform a reliable statistical analysis, with the ultimate goal of shedding light onto the physical processes driving BAL variability and onto the properties of the region where winds form and propagate.

In order to facilitate the identification of BAL troughs, we smooth our spectra by means of a three pixel-wide boxcar algorithm. We convert wavelengths into velocities through redshifts, and identify all the C {\footnotesize{IV}} BAL troughs present in each of the SDSS-I/II spectra: in order to be included in our sample, a trough must have a flux extending below 90\% of the normalized continuum level for a velocity span of $\Delta v \geq 2000 \mbox{ km s}^{-1}$, as per Eq. \ref{eqn:BI}. Such selection criteria return a sample of 1874 BAL troughs, detected in the spectra of the 1319 unique sources constituting our main sample.

Once we have identified the BAL troughs in the SDSS-I/II spectra, we inspect the corresponding regions in the BOSS spectrum/spectra associated with each of the sources in the main sample, to determine if BALs are still present in the same windows. We define a disappearance as when no absorption extends below $90\%$ of the normalized continuum level, or if a BAL transforms into a NAL ($\Delta v < 500$ km s$^{-1}$; this is a more conservative disappearance criterion than the $BI = 0$) in the corresponding BOSS spectrum. Associating BALs when comparing two different spectra is not always trivial, since the troughs can shift with respect to each other \citep[e.g.,][]{nur13,grier16}; we assume there is mutual correspondence between two troughs if they cover wavelength ranges that overlap at least partially. In cases where we have more than two spectra for a QSO, we choose to use the latest SDSS-I/II spectrum where a BAL trough is visible, and the earliest BOSS spectrum where it disappears, thus probing the shortest accessible timescales and the fastest variability. 

A total of 105 BAL troughs detected in the SDSS-I/II spectra of 94 unique sources disappear in the corresponding BOSS spectra. However, some criterion assessing the significance of the observed BAL disappearances is necessary in order to minimize contamination from spurious disappearances. Following \citet{nur12}, we perform a two-sample ${\chi}^2$ test on the two sets of data points corresponding to the flux in each pair of wavelength windows where we observe a disappearance, and require the probability $P_{{\chi}^2}$ associated with the test to be $\leq 10^{-8}$ for the change in a trough to be unlikely due to a random occurrence; hence, if $P_{{\chi}^2}\leq 10^{-8}$, we can discard the null hypothesis and be confident that the observed disappearance is real.

The defined threshold returns a sample of 56 disappearing troughs, detected in the spectra of 52 different sources (hereafter, we refer to this as the $P_8$ sample). Nevertheless, a visual inspection of each of our disappearing BAL candidates revealed that a number of excluded disappearances may in fact be real, suggesting that, while returning a highly reliable sample, our threshold might be overconservative. To address this possibility, we select a second sample of disappearances that appear reliable on the basis of visual inspection; this sample corresponds to troughs with a probability $P_{{\chi}^2} \leq 10^{-4}$ for disappearances to be accidental. The new requirement returns 17 additional disappearing BALs observed in 16 different sources; we shall refer to the full sample of sources for which $P_{{\chi}^2} \leq 10^{-4}$ as the $P_4$ sample. This consists of 73 (56+17) disappearing BAL troughs detected in the spectra of 67 sources.\footnote{There is one source that belongs to both subsamples, as it exhibits two disappearing BALs with $P_{{\chi}^2} \leq 10^{-8}$ and one disappearing BAL with $10^{-8} \leq P_{{\chi}^2} \leq 10^{-4}$; this is why the sum of the sources in the $P_4$ sample is 67 instead of 68.} In what follows we will generally report the results of our analysis for the $P_4$ sample, but we will also discuss some relevant results concerning the $P_8$ sample; this approach also allows a proper comparison between our findings and those from \citet{nur12}. In Table \ref{tab:bal_dataset} we report numerical details about the main sample as well as the $P_4$ and $P_8$ samples. 

Even when disappearing BAL troughs belong to the $P_8$ sample, residual absorption may be present (e.g., NALs are not taken into account). In four extreme instances (IDs 623, 735, 919, and 1638 in the extended version of Fig. \ref{fig:spec_overplot}) we observe a trough in the BOSS spectrum, indicating that there is still absorption, but the trough is above 90\% of the normalized flux level and hence it is not detected as a BAL/mini-BAL. In order to record such instances, we identify by visual inspection a ``pristine'' sample, following \citet{nur12}: the sample consists of all the disappearances where no residual absorption is detected, and includes 30 out of the 73 BAL troughs in the $P_4$ sample.
 
\begin{table}[tb]
\renewcommand\arraystretch{1.5}
\caption[Numerical information about the main sample and the $P_4$ and $P_8$ samples]{\footnotesize Detailed information about the main sample of sources, the $P_4$ and $P_8$ samples, and the C {\scriptsize{IV}} BAL troughs in their spectra. The main sample consists of all the sources for which at least a pair of spectra (one from SDSS-I/II and one from BOSS) is available and where at least one C {\scriptsize{IV}} BAL trough is detected in the SDSS-I/II spectrum/spectra.}
\begin{center}
\small
\setlength{\tabcolsep}{3mm}
\begin{tabular}{c c}
\toprule
\emph{MAIN SAMPLE} & \\
\hline
\vspace{0.2cm}
SDSS-I/II spectra & 1543\\
BOSS spectra & 1654\\
SDSS-I/II spectra exhibiting C {\scriptsize{IV}} BAL troughs & 1319\\
C {\scriptsize{IV}} BAL troughs detected in SDSS-I/II spectra & 1874\\
\midrule
\emph{$P_4$ SAMPLE} & \\
\hline
Sources with a disappearing BAL trough & 67\\
Disappearing BAL troughs in BOSS spectra & 73\\
\vspace{-0.15cm}
Sources belonging to the pristine sample & 4/30\\
($10^{-8} \leq P_{{\chi}^2} \leq 10^{-4}$) & \\
\midrule
\emph{$P_8$ SAMPLE} & \\
\hline
Sources with a disappearing BAL trough & 52\\
Disappearing BAL troughs in BOSS spectra & 56\\
Sources belonging to the pristine sample & 26/30\\
\bottomrule
\end{tabular}
\end{center}
\label{tab:bal_dataset}
\end{table}

Figure \ref{fig:spec_overplot} presents each pair of SDSS-I/II and BOSS spectra from the $P_4$ sample where C {\footnotesize{IV}} BAL disappearance is detected. Some QSOs with multiple BAL troughs have more than one disappearing BAL trough, though sometimes the additional BAL troughs do not disappear; we shall address this situation in Section \ref{section:coordination}.

\begin{figure}[tb]
 \centering
   {\includegraphics[width=10cm]{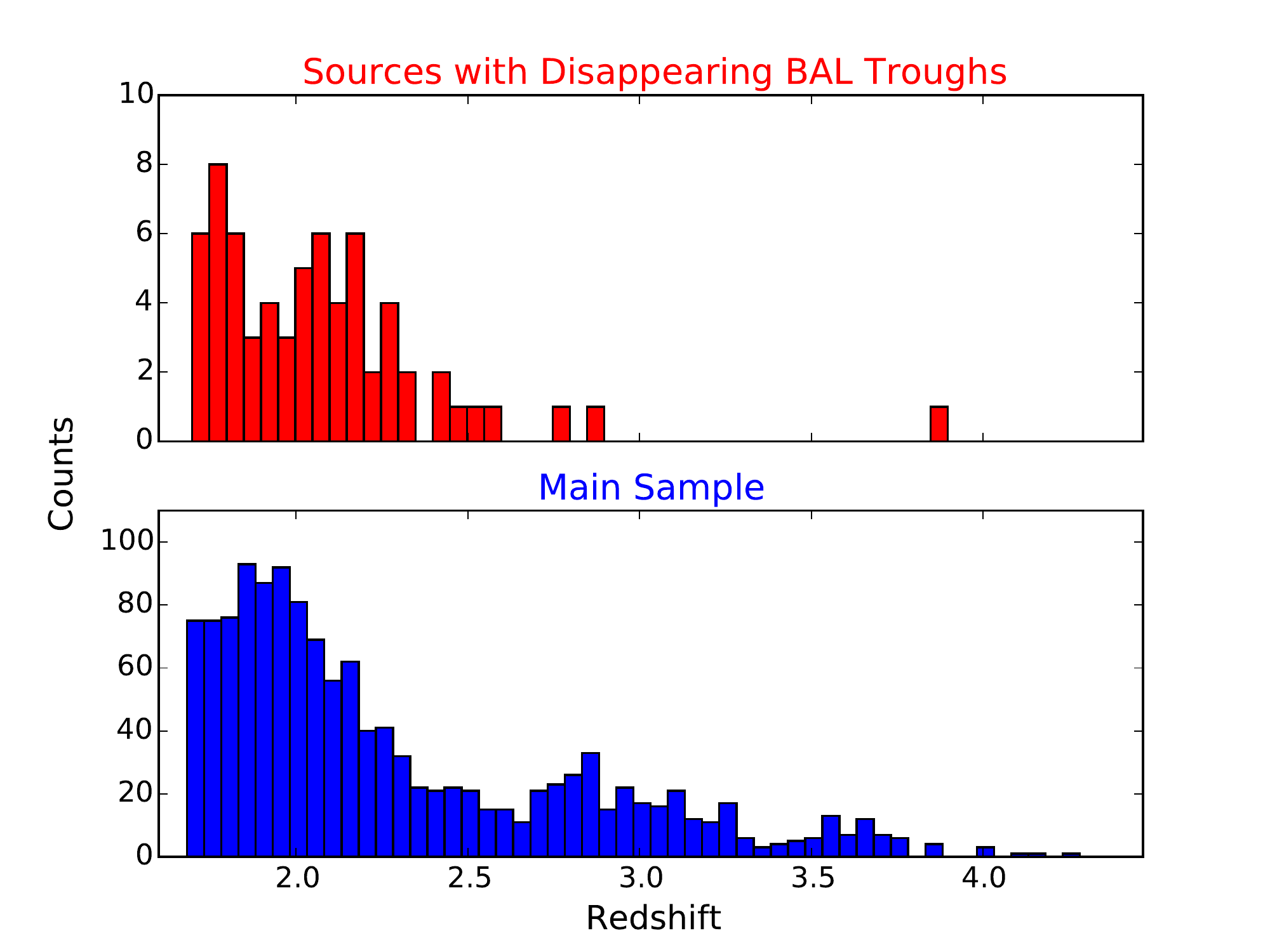}}	
   \caption{\footnotesize{Redshift distribution for the 67 sources in the $P_4$ sample (\emph{top}) and the 1319 sources in the main sample (\emph{bottom}). Redshifts are from \citet{hw}.}}\label{fig:redshift}
\end{figure}

\begin{figure}[tb]
 \centering
   {\includegraphics[width=10cm]{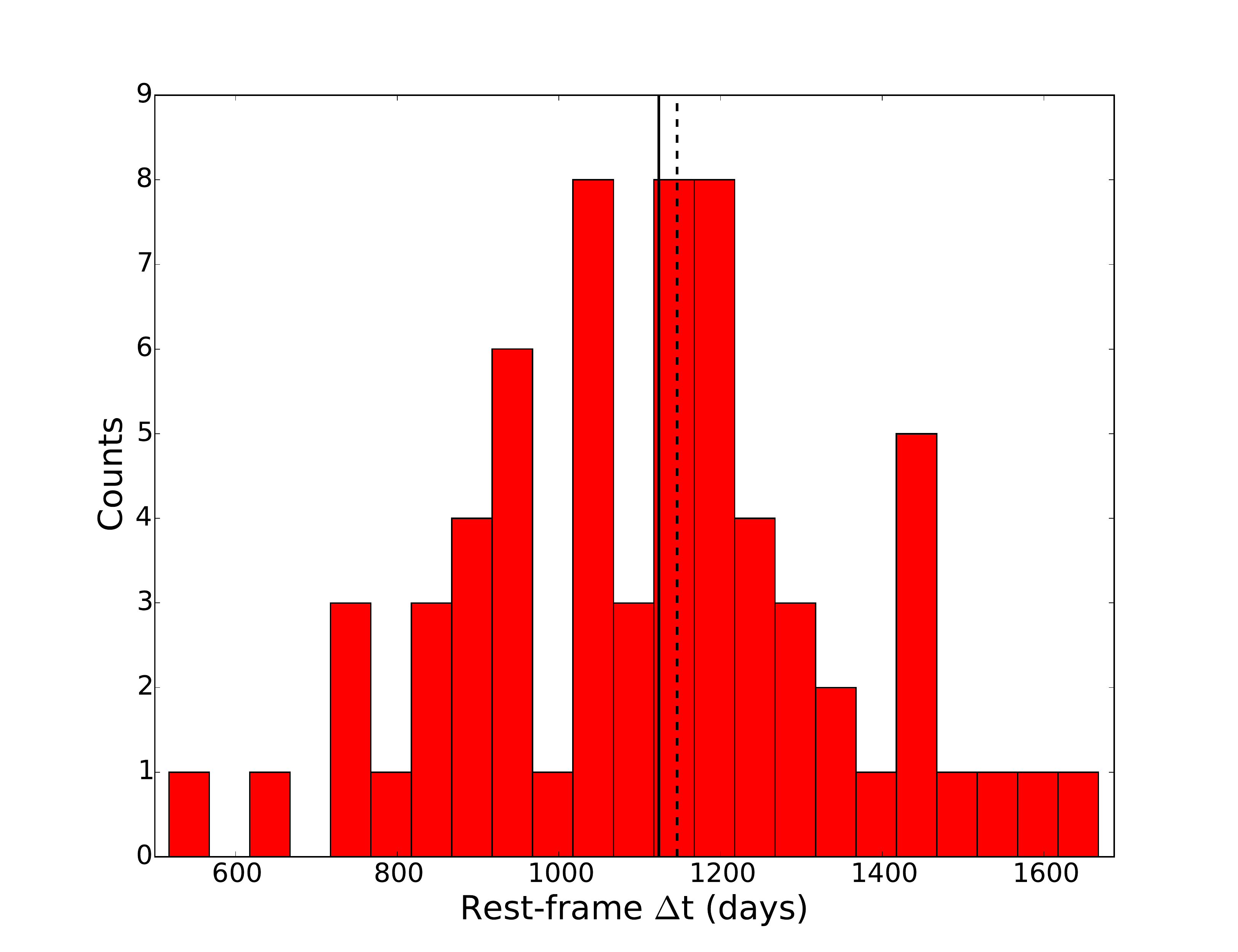}}
   \caption{\footnotesize{Distribution of the rest-frame time difference $\Delta t$ between the epoch pairs where disappearance is observed, for each source in the $P_4$ sample. The average (solid line) and median (dashed line) values for the time difference are shown. The typical rest-frame timescale for our sample is a few years.}}\label{fig:delta_t}
\end{figure}

Figure \ref{fig:redshift} displays the redshift distribution for all the sources in the main and $P_4$ samples: redshifts are in the range $1.68-4.27$. We note a lack of sources with disappearing BALs at $z \gtrsim 3$. Actually, the distribution of the rest-frame timespans for all the sources in the main sample with $z \gtrsim 3$ peaks at lower timespans ($\approx 750$ days) than the whole sample of sources with disappearing BALs (shown in Fig. \ref{fig:delta_t}), due to time dilation at higher redshifts. However, we do observe disappearing BALs on timescales of $\approx 750$ days at lower redshifts, so the lack of disappearing BALs for sources at $z \gtrsim 3$ does not appear to be a pure selection effect.
To assess the significance of our finding, we computed the fraction of sources with disappearing BALs for timescales shorter than the maximum time length sampled by our high-redshift sources, which turns out to be 3.3\%. Assuming that this occurrence rate is valid for our $z \gtrsim 3$ sources, we expect to find disappearing BALs in about five sources, while we find one. The likelihood of this happening by chance is 3\%; hence this result, although intriguing, is only marginally significant and requires a larger sample for proper investigation.

Figure \ref{fig:delta_t} shows the distribution of the rest-frame time difference $\Delta t$ between the two spectra in a pair where disappearance is observed for each source in the $P_4$ sample. 
The average $\Delta t$ is $\simeq 1123$ days, while the median $\Delta t$ is $\approx 1146$ days, and both correspond to a timescale of $\approx 3.1$ yr.

\subsection{Statistical properties of disappearing BAL troughs}
\label{subsection:bal_properties}
We have defined a main sample of 1319 sources where 1874 C {\footnotesize{IV}} BAL troughs have been detected, and we introduced the $P_4$ sample, consisting of 73 disappearing BAL troughs detected in the spectra of 67 sources; we also identified a more conservatively reliable sample of 56 disappearing BALs (the $P_8$ sample) observed in the spectra of 52 sources. 

On the basis of our findings, we can estimate the average lifetime of a BAL trough and of the BAL phase along our line of sight, to gain global insight into the BAL phenomenon over long timescales. We compute the fraction of disappearing BAL troughs $f_{\tiny{\mbox{disapp$_{_{P_4}}$}}}=73/1874=3.9_{-0.5}^{+0.5}\%$ and the fraction of QSOs exhibiting at least one disappearing BAL trough in their spectra $f_{\tiny{\mbox{QSO$_{_{P_4}}$}}}=67/1319=5.1_{-0.6}^{+0.7}\%$ (here and in the following, error bars on percentages are computed following \citealt{gehrels}, where approximated formulae for confidence limits are derived assuming Poisson and binomial statistics). The two fractions become $f_{\tiny{\mbox{disapp$_{_{P_8}}$}}}=56/1874=3.0_{-0.4}^{+0.5}\%$ and $f_{\tiny{\mbox{QSO$_{_{P_8}}$}}}=52/1319=3.9_{-0.5}^{+0.6}\%$, respectively, if we restrict our analysis to the $P_8$ sample.

The estimated disappearance frequency allows us to estimate the average rest-frame lifetime $\overline{t}_{\tiny{\mbox{trough$_{_{P_4}}$}}}$ of a BAL trough along our line of sight; we can roughly define it as the average value of the maximum time difference $\langle\Delta t_{\tiny{\mbox{max$_{_{P_4}}$}}}\rangle$ between two epochs in a pair in our main sample divided by the fraction of BAL troughs $f_{\tiny{\mbox{disapp$_{P_4}$}}}$ that disappear over such time. Since $\langle\Delta t_{\tiny{\mbox{max$_{_{P_4}}$}}}\rangle \approx 1144$ days, corresponding to $\approx 3.1$ yr, we obtain\footnote{Here and in what follows we adopt different notations to make a distinction between average quantities that we directly measure, e.g., $\langle\Delta t\rangle$, and average quantities that we derive, e.g., $\overline{t}$.} $\overline{t}_{\tiny{\mbox{trough$_{P_4}$}}}\approx \langle\Delta t_{\tiny{\mbox{max$_{P_4}$}}}\rangle/f_{\tiny{\mbox{disapp$_{P_4}$}}}=80_{-10}^{+10}$ yr. Limited to the BAL troughs in the $P_8$ sample, we obtain an average rest-frame lifetime $\overline{t}_{\tiny{\mbox{trough$_{P_8}$}}}\approx \langle\Delta t_{\tiny{\mbox{max$_{P_8}$}}}\rangle/f_{\tiny{\mbox{disapp$_{P_8}$}}}=104_{-14}^{+17}$ yr, the average value of the maximum time difference being $\langle\Delta t_{\tiny{\mbox{max$_{P_8}$}}}\rangle \approx 1142$ days.

Several works in the literature \citep[e.g.,][]{hall02,g09,nur12} have shown that, if a source is a BAL QSO, BALs originating from C {\footnotesize{IV}} transitions are generally present in its spectrum, and they are typically the strongest troughs. When all the C {\footnotesize{IV}} BAL troughs disappear from a spectrum, generally there are no remaining BALs, nor Ly$\alpha$ BALs\footnote{If no BALs corresponding to high-ionization transitions are observed, lower ionization BALs will generally not be observed as well.}, hence the source becomes a non-BAL QSO. We inspected the spectra in our $P_4$ sample and found that 30 sources change into non-BAL QSOs when C {\footnotesize{IV}} BAL troughs disappear; the rest of the sources exhibit additional non-disappearing C {\footnotesize{IV}} BAL/mini-BAL troughs in their BOSS spectra. We derive the fraction of QSOs turning into non-BAL QSOs as the ratio of the number of objects transforming into non-BAL QSOs to the total number of objects in our sample, that is, $f_{\tiny{\mbox{transform$_{_{P_4}}$}}}=30/1319=2.3_{-0.4}^{+0.5}\%$.

Once we know this fraction, we can estimate the lifetime of the BAL phase in a QSO, which we can roughly define as the average of the maximum time difference between epochs in a pair (already used above) divided by the fraction of BAL QSOs that turn into non-BAL QSOs over that time range, i.e., $\overline{t}_{\tiny{\mbox{BAL}}} \approx \langle\Delta t_{\tiny{\mbox{max}}}\rangle/f_{\tiny{\mbox{transform}}}$. Again, this is the observed BAL lifetime, related to a BAL being observed along the line of sight, rather than the lifetime of the outflowing gas.
For the $P_4$ sample $\overline{t}_{\tiny{\mbox{BAL$_{_{P_4}}$}}} \approx 136_{-24}^{+30}$ yr. If we focus on the sources in the $P_8$ sample, the number of BAL QSOs turning into non-BAL QSOs reduces to 24 and the corresponding fraction becomes $f_{\tiny{\mbox{transform$_{_{P_8}}$}}}=24/1319=1.8_{-0.4}^{+0.5}\%$; the lifetime of the BAL phase along the line of sight is therefore $\overline{t}_{\tiny{\mbox{BAL$_{_{P_8}}$}}} \approx 174_{-39}^{+48}$ yr. 

When dealing with such estimates, one should keep in mind that, even though all the BALs can disappear from the spectrum of a source, other BALs can emerge at a later time, either in that same region or in a different one. In addition, the disappearing rate is dependent on monitoring duration. Longer monitoring may mean more disappearing troughs (unless they reappear). As a consequence, the definition of ``BAL phase'', as well as the resulting $\overline{t}_{\tiny{\mbox{BAL}}}$, should be handled with caution.

\subsection{Velocity distributions}
\label{section:vel}
Relevant information about the BAL-trough population can be inferred from the analysis of BAL properties in terms of velocity. In Section \ref{section:intro} we introduced the maximum and minimum velocity of a BAL trough, $v_{\mbox{\tiny{max}}}$ and $v_{\mbox{\tiny{min}}}$, the velocity difference $\Delta v$, and the central velocity $v_{\mbox{\tiny{c}}}$. 

\begin{figure*}[htb]
 \centering
\subfigure
   {\includegraphics[width=8cm]{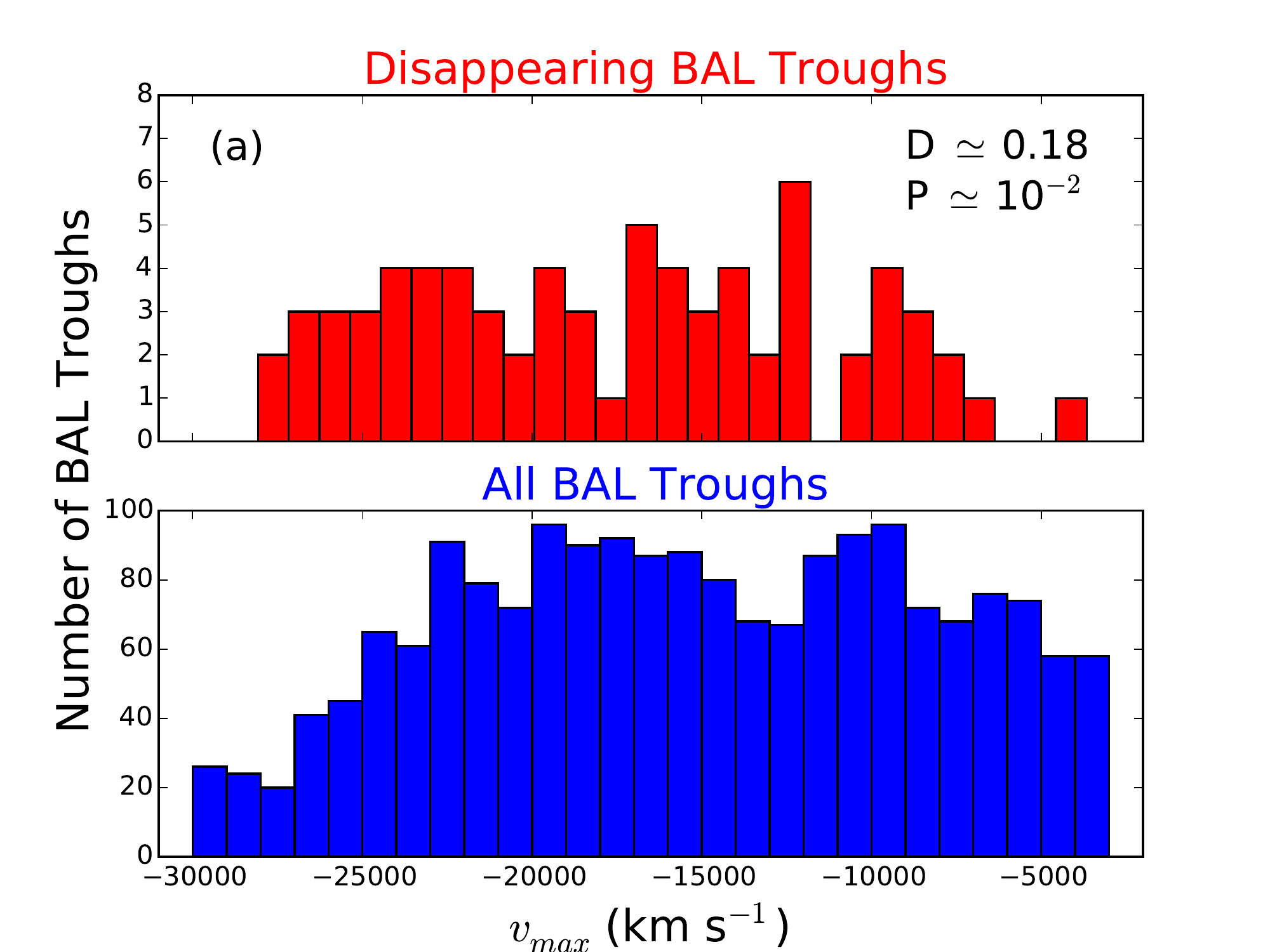}}		
\subfigure
   {\includegraphics[width=8cm]{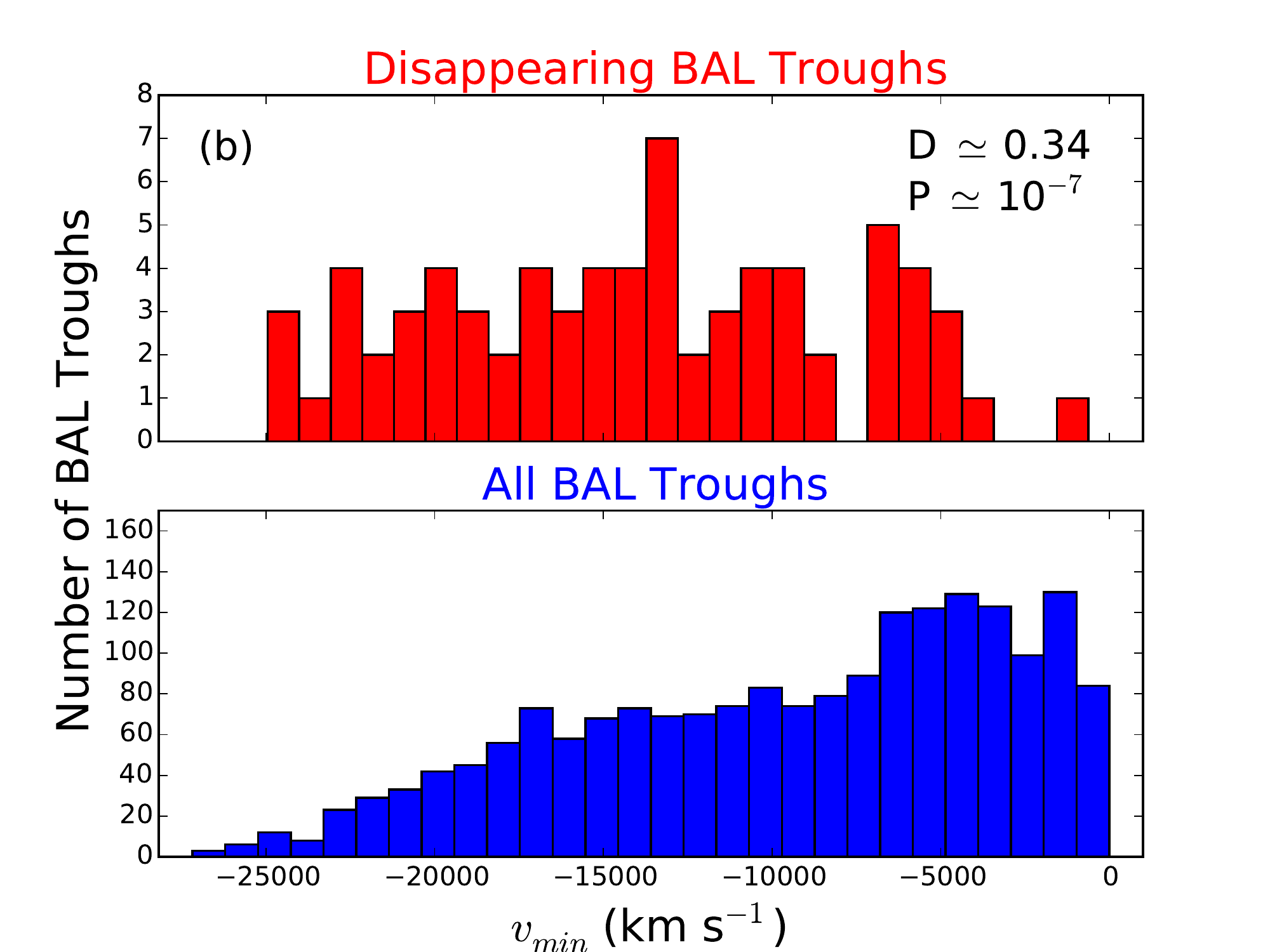}}		
\subfigure
   {\includegraphics[width=8cm]{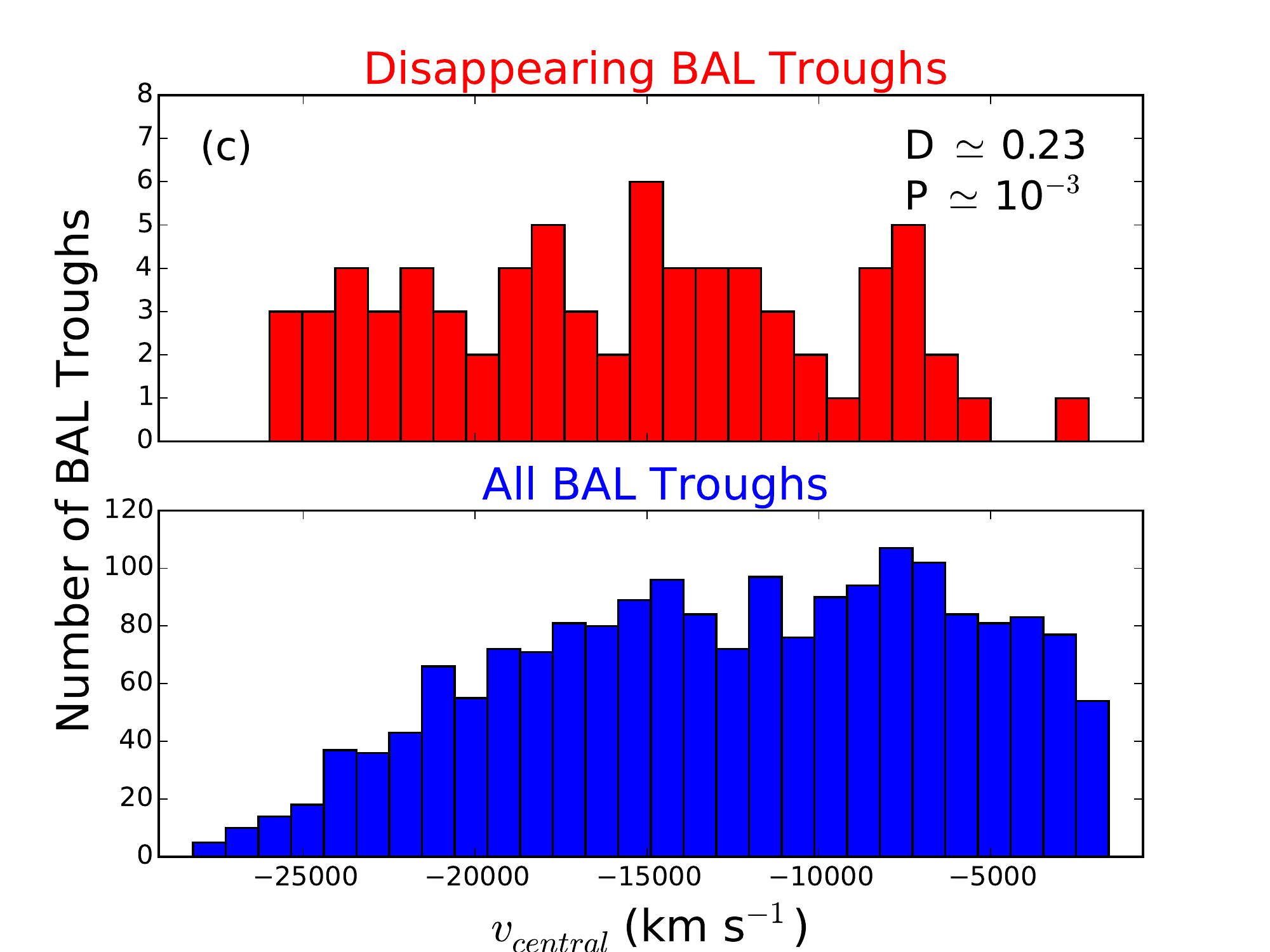}}		
\subfigure
   {\includegraphics[width=8cm]{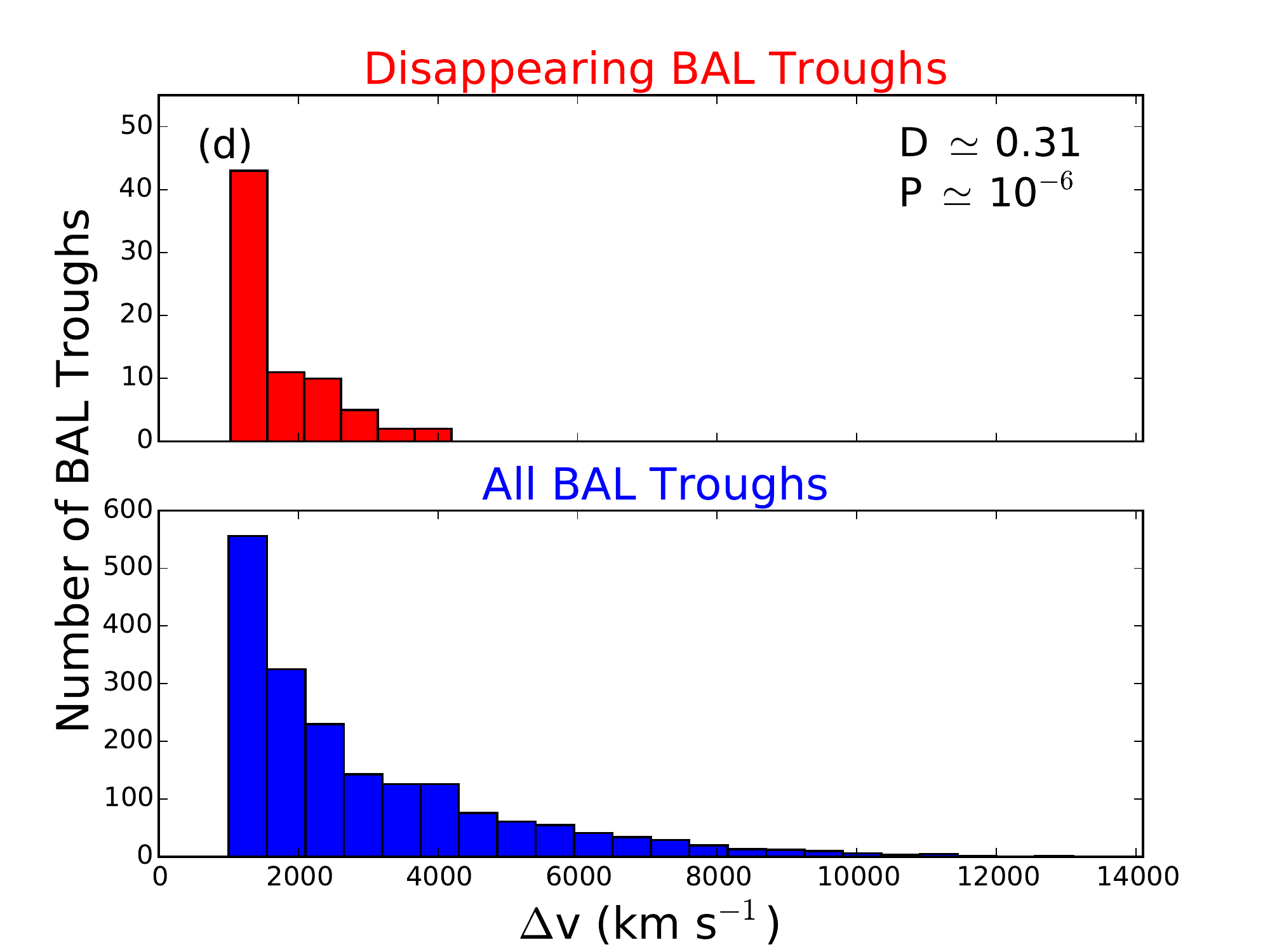}}		
   \caption{\footnotesize{Maximum observed velocity $v_{\mbox{\tiny{max}}}$ (\emph{a}), minimum observed velocity $v_{\mbox{\tiny{min}}}$ (\emph{b}), central velocity $v_{\mbox{\tiny{c}}}$ (\emph{c}), and BAL width $\Delta v$ (\emph{d}) distributions for the $P_4$ sample (upper histogram in each panel) and for the main sample of sources with BAL troughs (lower histogram in each panel). Results of the K-S test performed on each pair of cumulative distributions are reported in each panel: \emph{D} is the maximum distance between the two cumulative distributions, and \emph{P} is the probability of obtaining a higher \emph{D} value assuming that the two datasets are drawn from the same distribution function.}}\label{fig:v_hists}
   \end{figure*}

Figure \ref{fig:v_hists} compares the population of sources in the $P_4$ sample to the sources in the main sample. Specifically, in panels (\emph{a}) and (\emph{b}) the $v_{\mbox{\tiny{max}}}$ and $v_{\mbox{\tiny{min}}}$ distributions for both populations are shown, respectively. We perform a Kolmogorov-Smirnov (K-S) test on each pair of distributions (full results are reported in the various panels) in order to assess the probability of consistency between the two datasets. As for $v_{\mbox{\tiny{max}}}$, we find that the maximum distance between the two cumulative distributions is $D \approx 0.18$, and the probability to obtain a higher value for $D$ assuming that the two datasets are drawn from the same distribution function is $P \sim 10^{-2}$; as a consequence, we cannot state that the two distributions are inconsistent. 

Different results are obtained with the $v_{\mbox{\tiny{min}}}$ distributions, as it is apparent that the bulk of the minimum velocities for the $P_4$ sample is clustered around higher values when compared to the distribution derived for the main sample. In this case we measure a maximum distance $D \approx 0.34$ and a probability $P \sim 10^{-7}$, indicating that consistency is unlikely. This result gives confidence that our decision to exclude from our analysis the BALs entirely confined in the velocity range $-3000 - 0$ km s$^{-1}$ (see Section \ref{section:bal_sample}) does not affect significantly our results, since the $v_{\mbox{\tiny{min}}}$ distributions suggest that disappearing BALs are generally characterized by high values of $v_{\mbox{\tiny{min}}}$, and we are thus unlikely to have removed a significant number of objects from our sample.

In panel (\emph{c}) the central velocity distributions for the two samples are displayed: they resemble the distributions from panel (\emph{b}), and the disappearing BAL troughs have a higher central velocity than the BALs in the main sample. The probability in this case is higher than in panel (\emph{b}) because the central velocity is affected not only by $v_{\mbox{\tiny{min}}}$, but also by $v_{\mbox{\tiny{max}}}$ values. In panel (\emph{d}) the velocity difference distributions are compared; the disappearing BAL troughs are generally narrower than the ones in the main sample. 
\begin{figure*}[ht]
\vspace{-0.2cm}
 \centering
\vspace{-0.1cm}
\subfigure
   {\includegraphics[width=14.5cm]{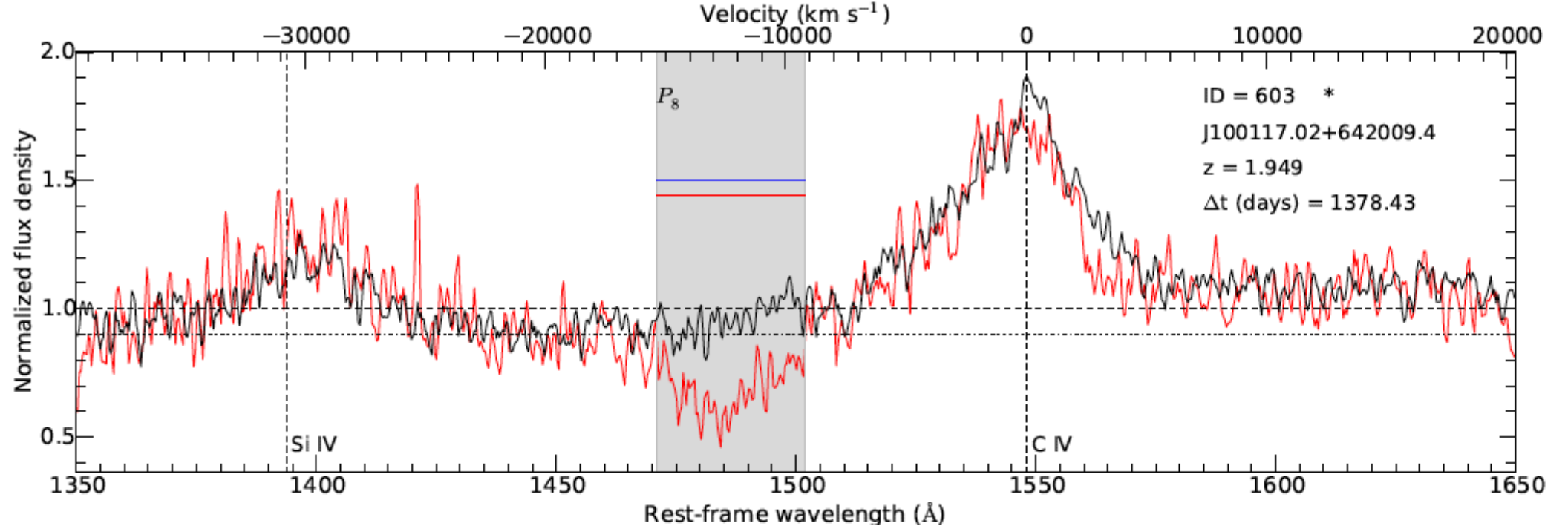}}
\vspace{-0.1cm}
\subfigure
   {\includegraphics[width=14.5cm]{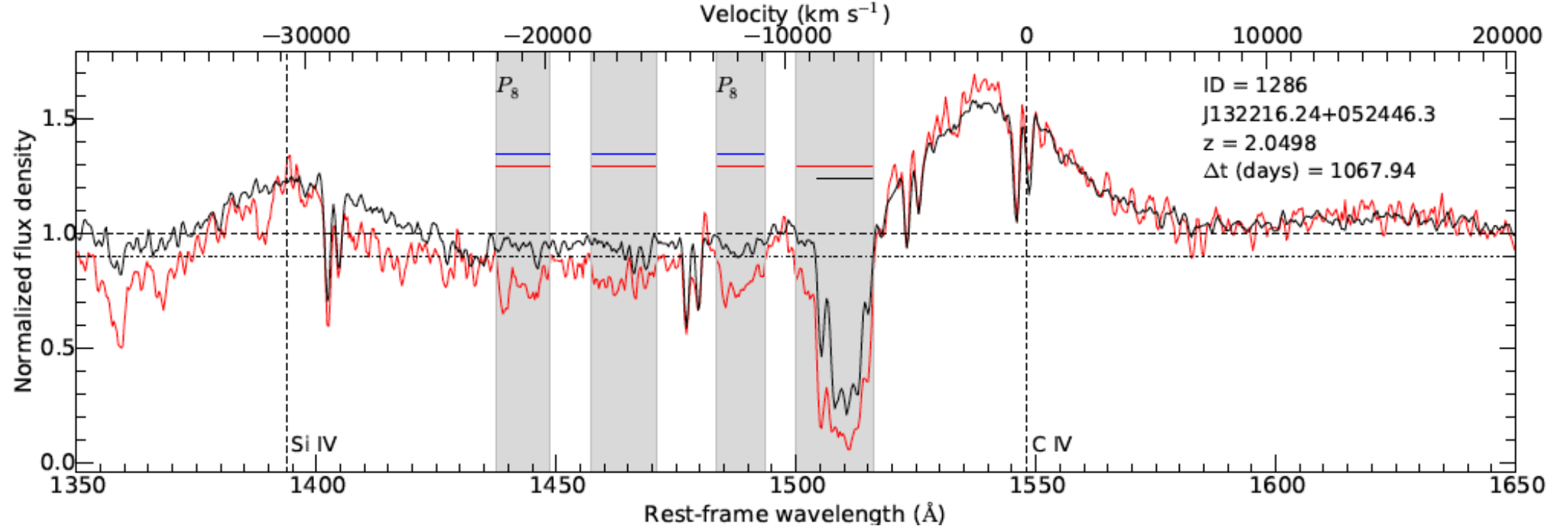}}
\vspace{-0.1cm}
\subfigure
   {\includegraphics[width=14.5cm]{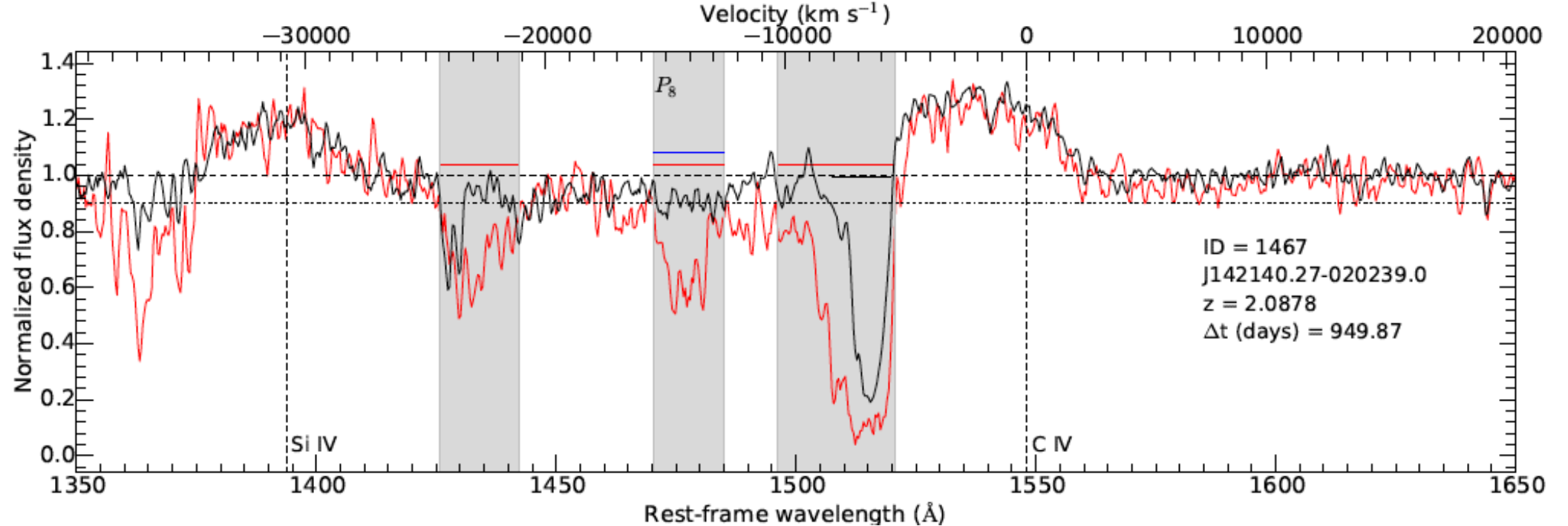}}
\vspace{-0.1cm}
\subfigure
   {\includegraphics[width=14.5cm]{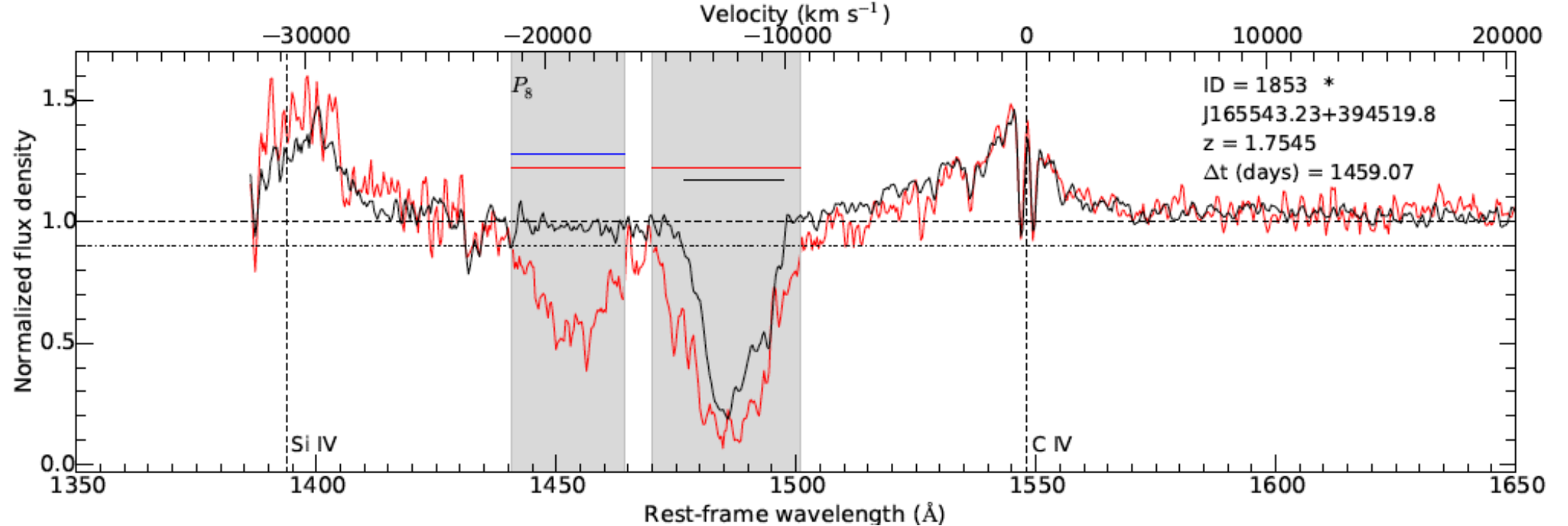}}
   \caption{\footnotesize{SDSS-I/II (red) and BOSS (black) spectra showing disappearing BAL troughs. The top right of each panel reports the ID in our catalog, the SDSS ID, the redshift, and the rest-frame time difference between the two plotted spectra; the presence of a star next to the source ID identifies the BAL troughs belonging to the pristine sample. Plots are limited to the wavelength range $1350 - 1650$ $\AA$ for a better visualization of the window where C {\scriptsize{IV}} BAL disappearance can be observed. Rest-frame wavelength (bottom) and velocity (top) are reported on the horizontal axes, while the normalized flux density is shown on the vertical axis. The horizontal dashed line represents the level where normalized flux density equals unity, while the dash-dot line indicates the 0.9 level for the normalized flux density; our analysis concerns BAL troughs extending below this threshold. The two vertical dashed lines represent the rest-frame wavelengths corresponding to the Si {\scriptsize{IV}} and C {\scriptsize{IV}} emission lines (1394 \AA$\mbox{ }$and 1549 \AA, respectively). Red and black horizontal lines identify BAL troughs in the SDSS-I/II spectra and BOSS spectra, respectively, while blue bars mark SDSS-I/II BAL troughs that disappear in BOSS spectra. The lack of either black lines or blue bars in a BOSS spectrum in correspondence of a BAL trough in the SDSS-I/II spectrum means that the BAL trough has turned into one or more mini-BALs, or that the disappearance is not considered reliable ($P_{{\chi}^2}>10^{-4}$). The regions corresponding to SDSS-I/II BAL troughs are shaded for better visualization. A ``$P_8$'' above a shaded area marks disappearing BAL troughs belonging to the $P_8$ sample. In some cases, BAL troughs outside the velocity range $-30000 - 0$ km s$^{-1}$ are apparent, but they are not taken into account in the present analysis (see Section \ref{section:bal_sample}). An extended version of this figure, including all the sources in our $P_4$ sample, is available in the online journal.}}\label{fig:spec_overplot}
\end{figure*}

\subsection{Equivalent widths and coordination in BAL variability}
\label{section:coordination}
In Section \ref{section:disapp} we mentioned that, in some cases, a spectrum of a ``disappearing'' BAL exhibits more than one C {\footnotesize{IV}} BAL trough, and not all disappear. In these cases we find BAL troughs in the earlier spectrum that are still BALs in the later spectrum, or BALs that turn into one or more mini-BALs; in this last case, on the basis of the definition of disappearance we introduced in Section \ref{section:intro}, we do not state that a BAL disappears. It can also happen that, regardless the number of disappearing/non-disappearing BALs, other BALs emerge in the later spectrum. Some of the listed instances were shown in Fig. \ref{fig:spec_overplot}. The presence of additional non-disappearing BAL troughs in spectra where disappearances are detected provides an opportunity for us to investigate the existence of a correlation in the variability of different BAL troughs when comparing two epochs of the same source.

Inspection of the 67 pairs of spectra corresponding to the sources in our $P_4$ sample reveals that there are 28 additional non-disappearing BALs in the spectra of 27 out of 67 sources. We choose not to take into account BALs turning into mini-BALs, and focus on BALs in the SDSS-I/II spectra that are still formally considered BALs in the corresponding BOSS spectra. Hereafter we shall refer to the subsample of the 28 additional non-disappearing BAL troughs (or, equivalently, to the subsample of the corresponding 27 QSOs) as the \emph{ND} sample. 

\begin{figure}[tb]
 \centering
 {\includegraphics[width=10cm]{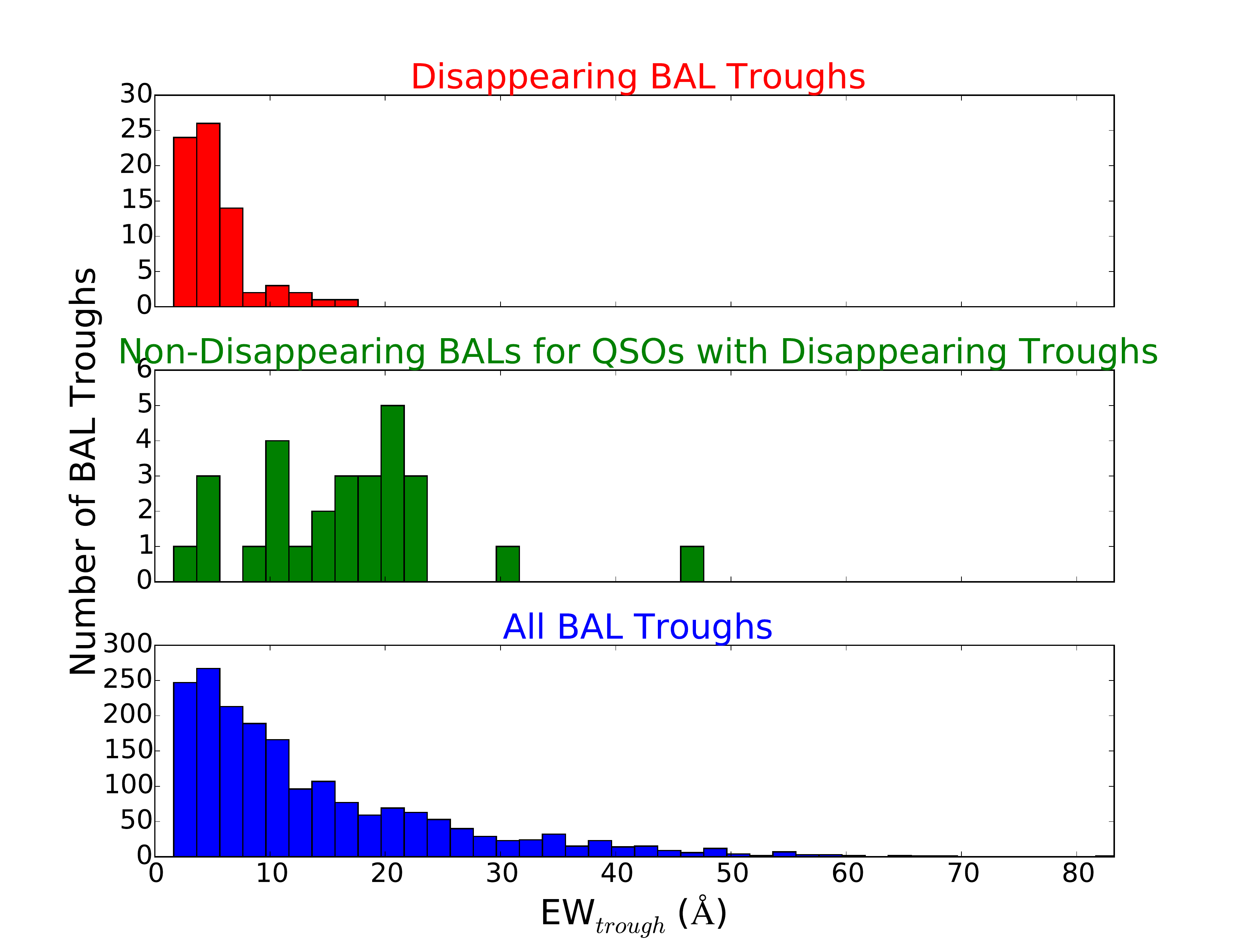}}	
 \caption[EW distributions]{\footnotesize{EW distributions for C {\scriptsize{IV}} BAL troughs in the $P_4$ sample (\emph{upper panel}), $ND$ sample (\emph{middle panel}), and main sample (\emph{lower panel}). EW measurements for each BAL are from the latest SDSS-I/II epoch.}}\label{fig:ew_hists}
\end{figure}

To further characterize our $P_4$ and \emph{ND} samples, we compute the EW of each BAL trough. Figure \ref{fig:ew_hists} displays the EW distributions for all the BAL troughs in the main sample and in the \emph{ND} and $P_4$ samples, in order to compare them. EW measurements are always performed in the latest SDSS-I/II spectrum in the case of QSOs where we observe a disappearance. In order to be consistent, they are performed in the latest SDSS-I/II epoch as well for the rest of the QSOs belonging to the main sample, which are used as a reference (see next figure). 

We again use a K-S test to compare the EW cumulative distributions. The probability of consistency for the main sample$-P_4$ sample pair is $P \sim 10^{-15}$ and the maximum distance is $D \approx 0.50$. Comparing the main sample to the \emph{ND} sample, produces $P \approx 0.006$ and a maximum distance $D \approx 0.32$ and so, in this last case, evidence for inconsistency is not as strong as in the previous one; nevertheless, the two distributions appear different from each other. The disappearing BAL troughs are generally characterized by low EW values, the highest one being $< 17$ \AA, while non-disappearing BAL troughs in the main sample typically reach much higher values ($\approx 80$ \AA) of EW.

\begin{figure}[tb]
 \centering
   {\includegraphics[width=10cm]{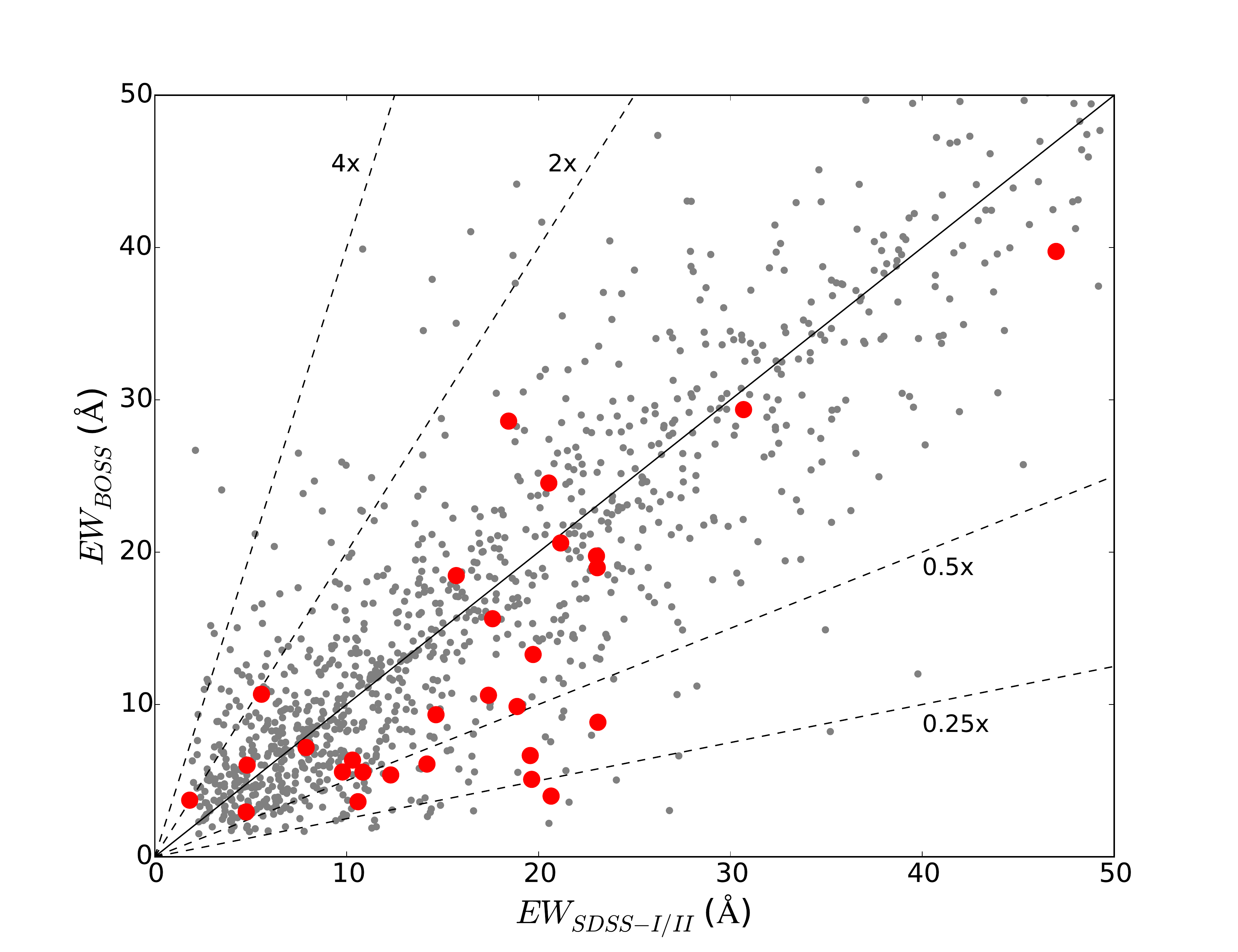}}		
   \caption{\footnotesize{EWs at two different epochs for the QSOs belonging to the \emph{ND} sample (large red dots). For each BAL trough, the two EW measurements are obtained from the same epoch pair where disappearance is detected. Small grey dots in the background show the EWs at two epochs (one from SDSS-I/II and one from BOSS) for all the sources in the main sample having non-disappearing BAL troughs. We limit our investigation to BAL troughs in SDSS-I/II spectra that correspond to BAL troughs in BOSS spectra, hence BALs turning into mini-BALs, or vice versa, are not taken into account. The solid line indicates where the EWs of the two compared epochs are equal, while the four dashed lines indicate where the EW of the BOSS epoch is four times, two times, half of, and a quarter of the EW in the SDSS-I/II epoch.}}\label{fig:ew_comparison}
\end{figure}

The comparison of the EWs of different BAL troughs in a pair of spectra allows investigation of the possible existence of coordination in the variability of such BALs. The EWs of the non-disappearing BAL troughs in our \emph{ND} sample are compared in Fig. \ref{fig:ew_comparison}; the two epochs are always the ones where we observe disappearing BALs (i.e., the latest SDSS-I/II epoch where we detect a BAL trough and the earliest BOSS epoch where that BAL is no longer detectable). All the non-disappearing BAL troughs detected in the spectra of QSOs belonging to the main sample are also presented as a reference; in this case, the two epochs chosen for the comparison are the latest among SDSS-I/II spectra and the earliest among BOSS spectra. 

The figure clearly shows that the distribution of the main-sample BALs is roughly symmetrical on the two sides of the bisector; this behavior indicates the absence of a dominant trend: BALs can become stronger as well as weaker. Conversely, there are 22 out of 28 ($79_{-17}^{+21}\%$) BAL troughs from the \emph{ND} sample that weaken over time: when there is more than one BAL trough and one of them disappears, in 79\% of the instances the EW of the remaining BALs decreases over time as well.

\begin{figure}[h!]
 \centering
   {\includegraphics[width=10cm]{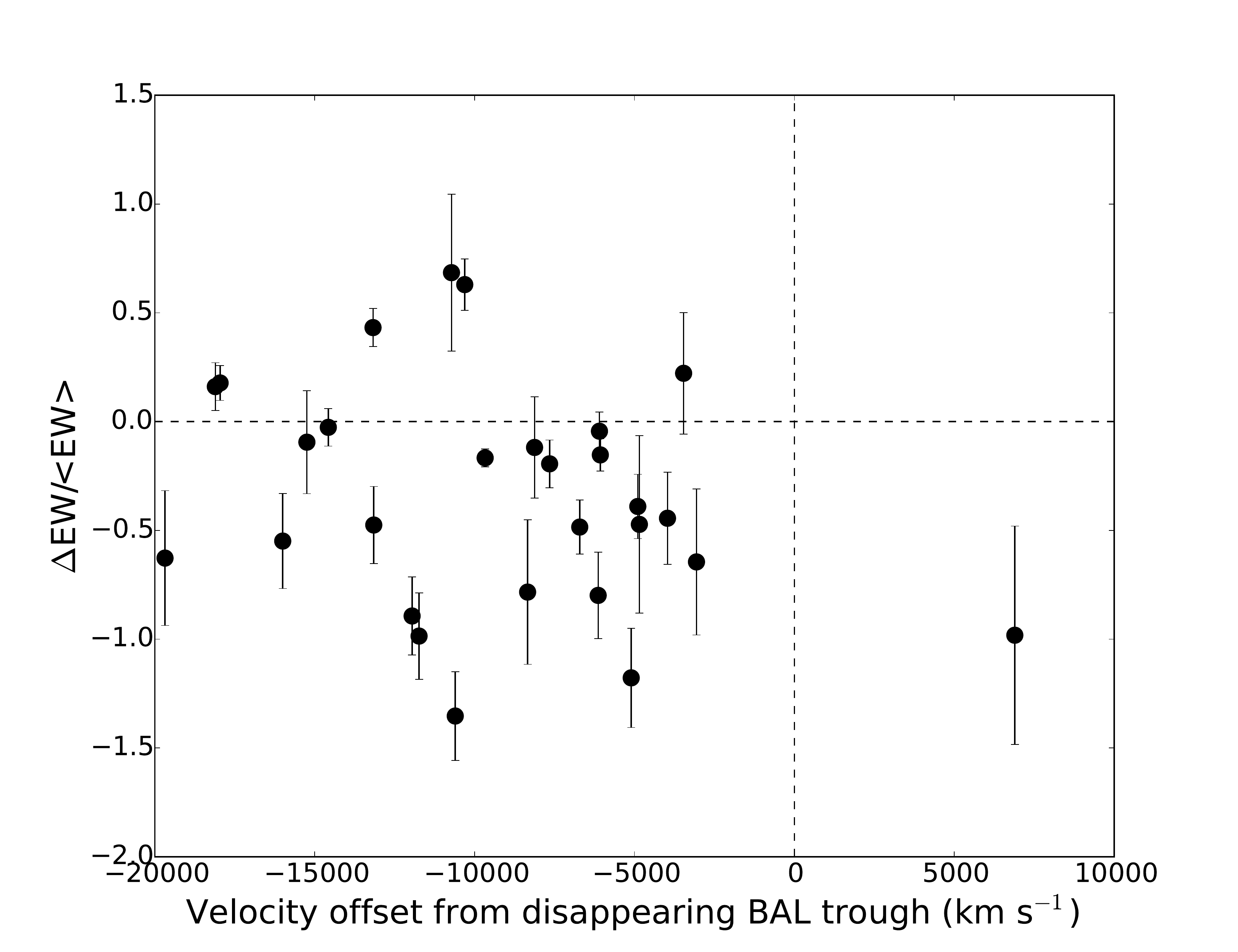}}
   \caption{\footnotesize{Fractional EW variations for the 28 non-disappearing BAL troughs in the \emph{ND} sample. The two epochs used for each trough are the same as in Fig. \ref{fig:ew_comparison}. The horizontal axis shows the offset of the central velocity $v_{\tiny{\mbox{c}_{_{\tiny{ND}}}}}$ of each non-disappearing trough with respect to the $v_{\mbox{\tiny{c}}}$ of the corresponding disappearing trough. The central velocity for a trough is defined as $v_{c} = (v_{max}+v_{min})/2$; error bars are computed propagating the errors on the EWs in the two epochs. The horizontal and the vertical dashed lines represent the zero levels for each axis, and are plotted for a better identification of the various regions in the diagram.}}\label{fig:frac_ew_var}
\end{figure}

As a further test, we analyze the fractional EW variation for the non-disappearing BAL troughs in the \emph{ND} sample: we define this quantity as $\Delta$EW$/\langle$EW$\rangle$, where $\Delta$EW $=$ EW$_{\mbox{\tiny{BOSS}}}-$EW$_{\mbox{\tiny{SDSS-I/II}}}$ and the average $\langle$EW$\rangle$ is computed over the two epochs. In Fig. \ref{fig:frac_ew_var} the fractional EW variation is presented as a function of the offset $v_{\mbox{\tiny{c}}} - v_{\tiny{\mbox{c}_{_{\tiny{ND}}}}}$ between the central velocity $v_{\tiny{\mbox{c}_{_{\tiny{ND}}}}}$ of the non-disappearing BAL and the central velocity $v_{\mbox{\tiny{c}}}$ of the disappearing BAL detected in the same pair of epochs; where there is more than one disappearing BAL, the offset is computed with respect to the one having the highest central velocity. In 27 out of 28 (96\%) of the cases, the BAL trough that disappears is the one with the highest central velocity. Moreover, the already mentioned weakening trend is apparent: only six of the non-disappearing BALs (21\% of 28) are stronger in the BOSS epoch than in the SDSS-I/II epoch. The BAL troughs with a positive velocity offset (i.e., those with $v_{\tiny{\mbox{c}_{\tiny{ND}}}} < v_{\mbox{\tiny{c}}}$) generally weaken. The weakening trend is also observed in the BALs with the largest velocity offsets; all of these results demonstrate the existence of coordination in BAL-trough variability and also suggests this behavior is a persistent phenomenon. 

\subsection{Comparison to results from Filiz Ak et al. (2012)}
\label{section:nur}
In Section \ref{section:intro} we mentioned that a subset of the spectra that we analyze in our work was also inspected by \citet{nur12}; a comparison of the findings is presented here. 

First, we cross-match our main sample to the sample of QSOs examined in \citet{nur12}, where observations have MJD $\leq 55811$, and the sample consists of 582 sources where 925 C {\footnotesize{IV}} BAL troughs are identified. The corresponding sample of disappearing BALs consists of 21 troughs detected in the spectra of 19 QSOs (hereafter, the $F12$ sample). The cross-match of our main sample with the $F12$ sample returns 558 out of 582 sources; here we list the reasons why the remaining 24 objects are not in our main sample:
\begin{itemize}
\item[--] seven sources are excluded since their SDSS-I/II spectra exhibit C {\footnotesize{IV}} BAL troughs outside the velocity range of interest of this paper (all but one have $v_{\mbox{\tiny{max}}} > -3000$ km s$^{-1}$, while the other has $v_{\mbox{\tiny{min}}} < -30000$ km s$^{-1}$);
\item[--] seven sources are excluded since their SDSS-I/II spectra do not show C {\footnotesize{IV}} BAL troughs in the velocity range of interest of this paper. Nonetheless, a careful inspection of the spectra at issue reveals that all of them possess at least one mini-BAL trough in the velocity range of interest; each mini-BAL has a width $\Delta v > 1500$ km s$^{-1}$ and, in particular, half of them have $\Delta v > 1920$ km s$^{-1}$. This suggests that we do not detect the expected BALs due to slight differences in the spectrum fitting/normalization in the two works;
\item[--] six sources are excluded as no SDSS-I/II (one instance) or BOSS spectrum (five instances) is available for them. Their spectra were available at the time the work by \citet{nur12} was ongoing, but they were later excluded from the SDSS archive as the corresponding sources have very close neighbors, and this caused mismatching;
\item[--] two sources are excluded because of problems in fitting the continuum of their SDSS-I/II spectra;
\item two sources are excluded as they belong to the sample of ten sources mentioned in Section \ref{section:bal_sample}, with spectrum pairs exhibiting non-overlapping continua due to a vertical shift.
\end{itemize}

\begin{figure*}[htb]
 \centering
\addtolength{\subfigcapskip}{-0.5cm}
\vspace{-0.15cm}
\subfigure
   {\includegraphics[width=12.5cm]{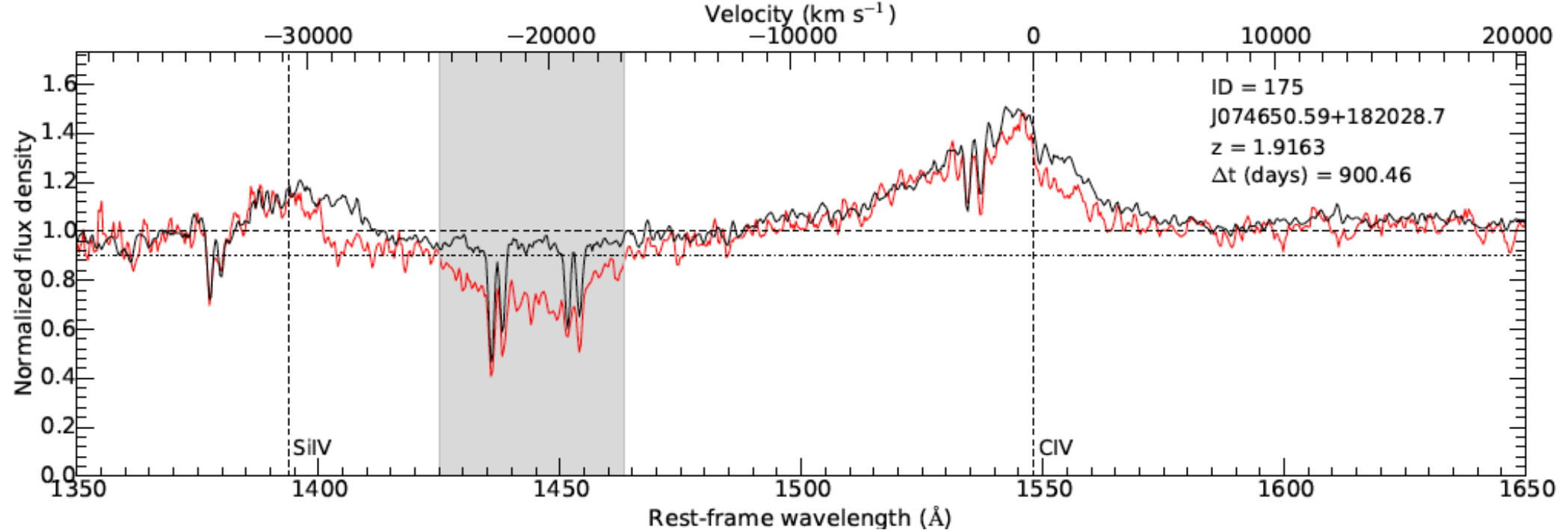}}
\vspace{-0.15cm}
\subfigure
   {\includegraphics[width=12.5cm]{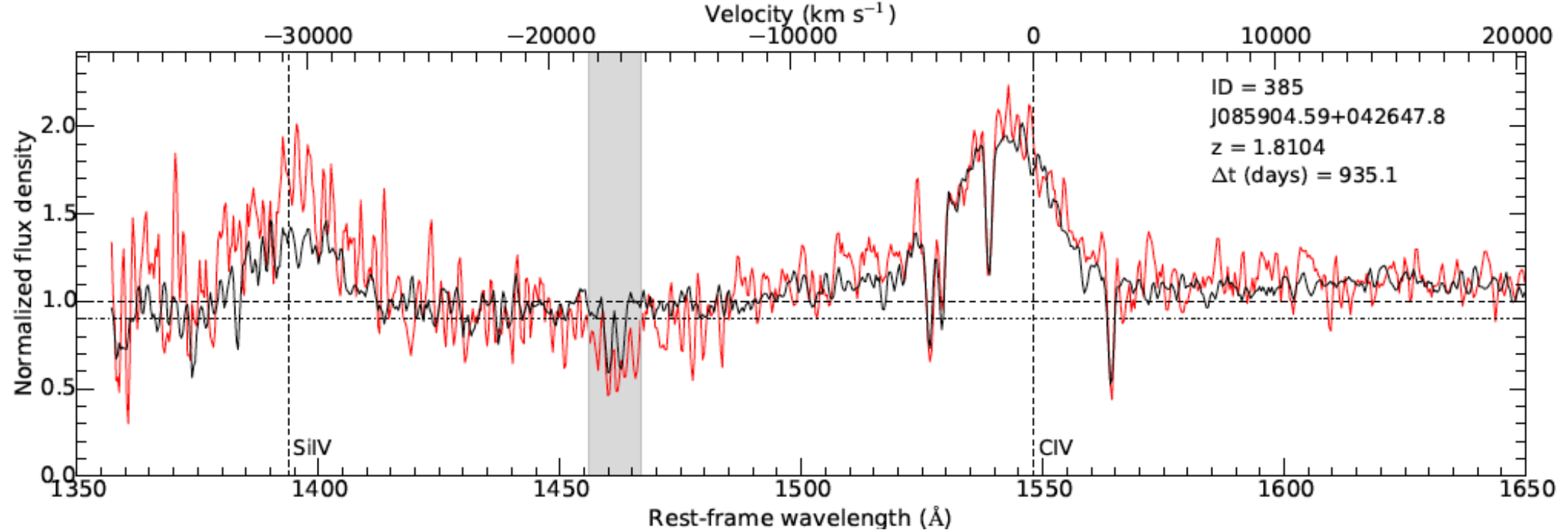}}
\vspace{-0.15cm}
\subfigure
   {\includegraphics[width=12.5cm]{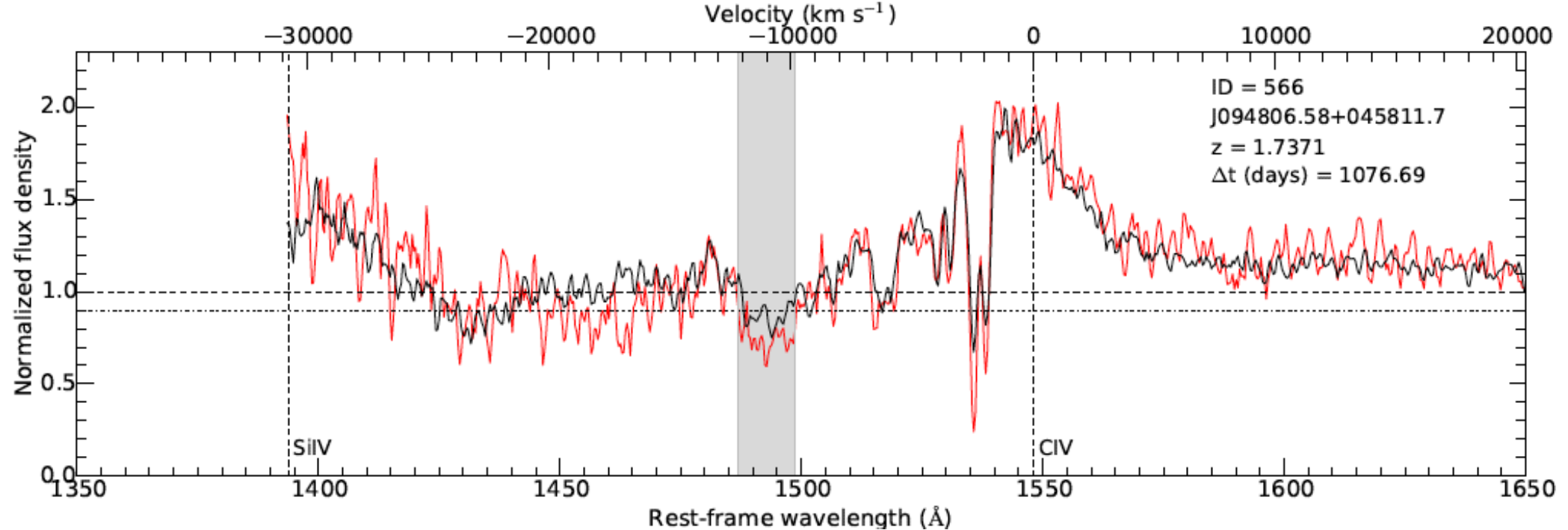}}
\vspace{-0.15cm}
\subfigure
   {\includegraphics[width=12.5cm]{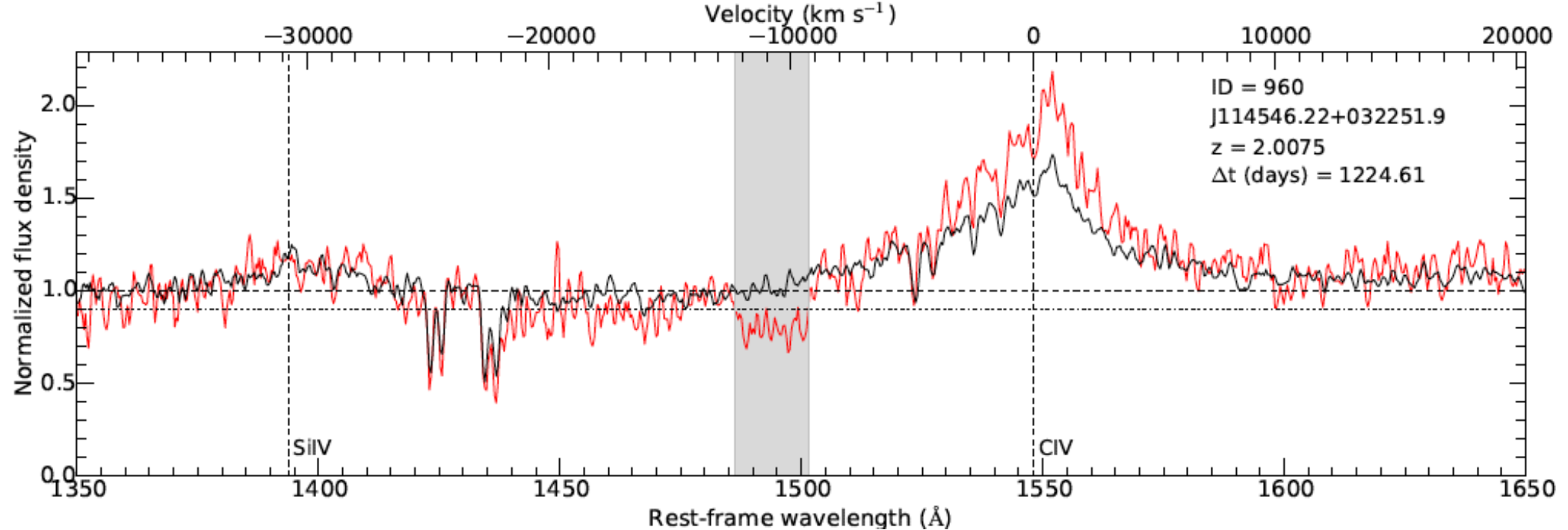}}
\vspace{-0.15cm}
\subfigure
   {\includegraphics[width=12.5cm]{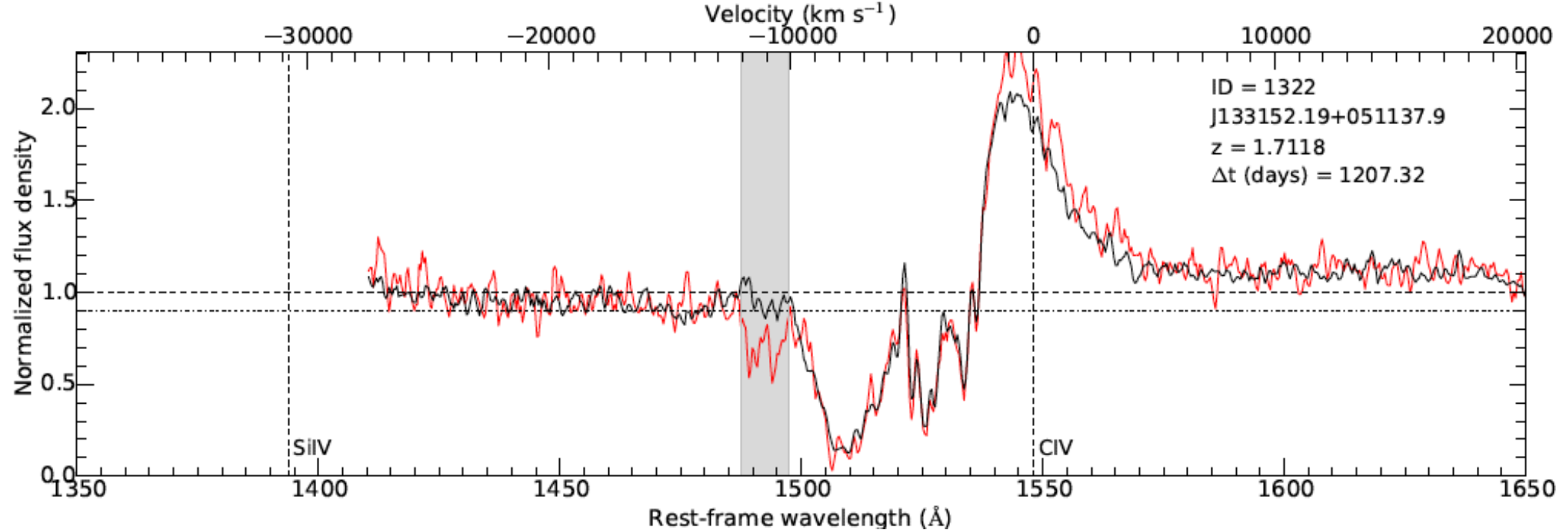}}
   \caption{\footnotesize{The five SDSS-I/II (red) and BOSS (black) spectra where disappearing BAL troughs were detected by \citet{nur12}, but not in the present work. The top right of each panel contains the ID of the corresponding source from our catalog, the SDSS ID, the redshift, and the rest-frame time difference between the two spectra represented. Plots are limited to the wavelength range $1350 - 1650$ $\AA$ for a better visualization of the window where C {\scriptsize {IV}} BAL disappearance can be observed. Rest-frame wavelength (bottom) and velocity (top) are reported on the horizontal axes, while the normalized flux density is shown on the vertical axis. The horizontal dashed line represents the level where normalized flux density equals unity, while the dash-dot line indicates the 0.9 level for the normalized flux density; our analysis concerns BAL troughs extending below this threshold. The two vertical dashed lines represent the rest-frame wavelengths corresponding to the Si {\scriptsize {IV}} and C {\scriptsize {IV}} emission lines (1394 \AA$\mbox{ }$and 1549 \AA, respectively). In the case of QSO J114546.22+032251.9, there are two peaks --and not just one-- in the SDSS-I/II spectrum crossing the threshold upwards, so three adjacent troughs comprise the BAL; nevertheless, the first two corresponding to higher velocities are sufficient to make a BAL trough if their widths are summed.}}\label{fig:fa_bals}
\end{figure*}

From the $P_4$ sample we retrieve 16 of the 21 disappearing BAL troughs constituting the $F12$ sample. A detailed analysis of the spectra of the five undetected disappearing BALs reveals that:
\begin{itemize}
\item[--] the BOSS spectrum of J074650.59+182028.7 exhibits two non-deblended NAL doublets in the wavelength region where the BAL from the SDSS-I/II spectrum is supposed to disappear; since they are not deblended, the two doublets appear as mini-BALs, and this is not classified as a disappearance;
\item[--] the SDSS-I/II spectrum of J133152.19+051137.9 contains a trough whose measured width is $< 1$ km s $^{-1}$ below our threshold defining BALs ($\Delta v \geq 2000$ km s$^{-1}$), and hence it cannot be considered as a BAL, technically;
\item[--] in the remaining three cases (J085904.59+042647.8, J094806.58+045811.7, J114546.22+032251.9) we do not detect any BAL troughs in the region indicated in \citet{nur12} in the corresponding SDSS-I/II epoch because of the presence of a narrow rise in the spectrum crossing the BAL-trough threshold (i.e., $90\%$ of the normalized continuum level) upwards; as a consequence, our interpretation is two adjacent troughs, each having a width $\Delta v < 2000$ km s$^{-1}$. The difference is likely caused by slight differences in the continuum fits between our sample and \cite{nur12}.\end{itemize}
Figure \ref{fig:fa_bals} presents the pairs of spectra corresponding to each of the five mentioned QSOs.

\cite{nur12} measure the fraction of disappearing BAL troughs to be $f_{\tiny{\mbox{disapp$_{_{F12}}$}}}=21/925=2.3_{-0.5}^{+0.6}\%$, while the fraction of QSOs showing at least one disappearing BAL trough in their spectra is $f_{\tiny{\mbox{QSO$_{_{F12}}$}}}=19/582=3.3_{-0.7}^{+0.9}\%$; both percentages are slightly lower than the ones for our $P_4$ sample, but are consistent with the percentages obtained from the analysis of our $P_8$ sample. The estimated average BAL-trough lifetime from \citet{nur12} is $\overline{t}_{\tiny{\mbox{trough$_{_{F12}}$}}}=109_{-22}^{+31}$ yr, while the BAL-phase duration is $\overline{t}_{\tiny{\mbox{BAL$_{_{F12}}$}}}=150_{-50}^{+60}$ yr: both are consistent with the estimates derived both from the $P_8$ and the $P_4$ samples. 

In Section \ref{section:disapp} we mentioned that, following \citet{nur12}, we defined a pristine sample consisting of all the disappearing C {\footnotesize{IV}} BAL troughs where we do not detect any residual absorption. In \citet{nur12} the pristine sample consists of 11 out of 21 BAL troughs. We classify as pristine seven out of these 11 BALs, while three of them belong to the instances where no BAL troughs are detected, shown in Fig. \ref{fig:fa_bals} (J074650.59+182028.7, J085904.59+042647.8, J094806.58+045811.7), and are therefore excluded from our analysis. We do not classify as pristine the remaining BAL trough (J155119.14+304019.8; see the extended version of Fig. \ref{fig:spec_overplot}).

\section{Summary and discussion}
\label{section:discussion}
The present work analyzed the disappearance of C {\footnotesize{IV}} BAL troughs in the largest sample of BAL QSOs investigated to date, produced by the SDSS-I/II/III surveys; our ultimate goal is a deeper understanding of the physics and structure of QSOs, to be possibly investigated in a future follow-up work. We selected a sample of QSOs exhibiting C {\footnotesize{IV}} BAL troughs in their spectra and performed a statistical analysis of the subsample of BALs that disappear with the aim of extending our knowledge of the physical processes driving BAL variability. Here follows a list of our main findings.
\begin{enumerate}[i.]
\item The main sample of QSOs consists of 1319 sources with 1874 detected C {\footnotesize{IV}} BAL troughs in their spectra; the sample of sources showing disappearing BALs consists of 67 QSOs, with 73 disappearing BALs (the $P_4$ sample). Such disappearances are observed over a rest-frame timescale of $0.28-4.9$ yr. The fraction of sources with disappearing BALs is $f_{\tiny{\mbox{QSO$_{_{P_4}}$}}}=5.1_{-0.6}^{+0.7}\%$, while the fraction of disappearing BALs is $f_{\tiny{\mbox{disapp$_{_{P_4}}$}}}=3.9_{-0.5}^{+0.5}\%$.\\
In Section \ref{section:bal_sample} we mentioned ten sources that we excluded from our analysis because the continua in each spectrum pair did not overlap. It is worth noting that, if we included these sources in our main sample, we would detect C {\footnotesize{IV}} BAL troughs in the SDSS-I/II spectra of the ten sources, and would find that in two instances these BALs disappear. This would lead to a fraction of sources with disappearing BALs $5.2_{-0.6}^{+0.7}\%$, which is perfectly consistent with the one we obtained. This confirms that our conservative choice to exclude the ten sources from our analysis had no significant impact on our results.
\item We estimated the average BAL lifetime --limited to the direction of our line of sight-- to be $\overline{t}_{\tiny{\mbox{trough$_{P_4}$}}}\approx 80_{-10}^{+10}$ yr. Some clarification is necessary: the disappearing BALs tend to be weak, having an EW $< 17$ \AA, as shown by their EW distribution in Fig. \ref{fig:ew_hists}. This result means that our lifetime estimates characterize low-EW BALs, but nothing can be stated about BALs with a higher EW, as they have not been observed to disappear. A possible explanation for BAL disappearance is disk rotation, with disappearances occurring because BALs move out of our line of sight, while the relevant absorbing material still exists physically. In this context, our BAL-lifetime estimate would correspond to the orbital time of the accretion disk at distances of $\approx 0.09$ pc, thus placing the origin of BAL absorption at larger radii than those reported in some literature \citep[e.g.,][]{murray}, but fairly consistent with some other works \citep[e.g.,][]{nur13}. 
\item Thirty of the BAL QSOs in our $P_4$ sample turn into non-BAL QSOs when BALs in their spectra disappear. We computed the corresponding fraction of transforming BAL QSOs, which is $f_{\tiny{\mbox{transform$_{_{P_4}}$}}}=2.3_{-0.4}^{+0.5}\%$, and estimated the average lifetime of the BAL phase in a QSO, $\overline{t}_{\tiny{\mbox{BAL$_{_{P_4}}$}}} \approx 136_{-24}^{+30}$ yr. Once again, the estimate is limited to what we can measure along our line of sight. 
\item We selected a more conservatively reliable sample of 56 disappearing BALs in 52 sources, namely the $P_8$ sample; the mentioned quantities for such a sample become, respectively: $f_{\tiny{\mbox{QSO$_{_{P_8}}$}}}=3.9_{-0.5}^{+0.6}\%$, $f_{\tiny{\mbox{disapp$_{_{P_8}}$}}}=3.0_{-0.4}^{+0.5}\%$, $\overline{t}_{\tiny{\mbox{trough$_{P_8}$}}}\approx104_{-14}^{+17}$ yr, $f_{\tiny{\mbox{transform$_{_{P_8}}$}}}=1.8_{-0.4}^{+0.5}\%$, and $\overline{t}_{\tiny{\mbox{BAL$_{_{P_8}}$}}} \approx 174_{-39}^{+48}$ yr.
\item The distributions of minimum velocity, central velocity, and velocity difference for our main sample and for the $P_4$ sample (see Section \ref{section:vel}), as well as the corresponding EW distributions (Section \ref{section:coordination}), show that the BAL troughs that disappear are generally narrow and characterized by a higher outflow velocity with respect to non-disappearing BALs. 
\item The analysis provided evidence for the existence of a coordination in the variability of multiple troughs corresponding to the same transition, as apparent from the spectra shown in Section \ref{section:coordination}: in spectra where more than one BAL is detected and one of them disappears, the other BALs weaken in 79\% of these cases, while in the main-sample population there is no dominant trend between strengthening and weakening. Also, in 96\% of the cases the disappearing BAL is the one with the highest outflow velocity. Coordination in variability persists even when the radial distances between the two BALs appear to be very large (central velocity offset up to $\approx 20000$ km s$^{-1}$). 
\item We compared our findings to the results from \citet{nur12}, where part of our sample of BAL QSOs was analyzed; all of these sources are reported in Table \ref{tab:comparison} to allow a straightforward comparison. It is apparent that all the results for our $P_8$ sample are consistent with the corresponding results from \citet{nur12}; if we focus on the $P_4$ sample, this statement still holds, except for the first two fractions reported in the table, which are slightly larger in our analysis, even when accounting for errors.
\end{enumerate}

A detailed list of the QSOs in our $P_4$ sample and a list of parameters of the disappearing BALs in the $P_4$ sample are reported in Table \ref{tab:disapp_bals} and Table \ref{tab:disapp_bals1}, respectively, at the end of this paper.
\begin{table*}[tb]
\renewcommand\arraystretch{1.5}
\caption{\footnotesize Comparison between the main numerical findings from the present work and \citet{nur12}.}
\label{tab:comparison}
\begin{center}
\footnotesize
\begin{tabular}{l c c c}
\toprule
\	& $P_4$ sample & $P_8$ sample & \citet{nur12} \\
Fraction of sources & $5.1^{+0.7}_{-0.6}\%$ & $3.9^{+0.6}_{-0.5}\%$ & $3.3^{+0.9}_{-0.7}\%$\\
\vspace{0.2cm}
with disappearing BAL troughs & (67/1319) & (52/1319) & (19/582)\\
Fraction of disappearing & $3.9^{+0.5}_{-0.5}\%$ & $3.0^{+0.5}_{-0.4}\%$ & $2.3^{+0.6}_{-0.5}\%$\\
\vspace{0.2cm}
BAL troughs & (73/1874) & (56/1874) & (21/925)\\
Average BAL-trough & & & \\
\vspace{0.2cm}
lifetime $\overline{t}_{\tiny{\mbox{trough}}}$ (yr) & $80_{-10}^{+10}$ & $104_{-14}^{+17}$ & $109_{-22}^{+31}$\\
Fraction of BAL QSOs & $2.3_{-0.4}^{+0.5}\%$ & $1.8_{-0.4}^{+0.5}\%$ & $1.7_{-0.5}^{+0.7}\%$\\
\vspace{0.2cm}
that turn into non-BAL QSOs & (30/1319) & (24/1319) & (10/582)\\
Average BAL-phase lifetime $\overline{t}_{\tiny{\mbox{BAL}}}$ (yr) & $136_{-24}^{+30}$ & $174_{-39}^{+48}$ & $150_{-50}^{+60}$\\\bottomrule
\end{tabular}
\end{center}
\end{table*}

BALs are thought to form because of outflowing winds originating in the proximity of the central SMBH, and the existence of a coordination in their variability sheds light onto the possible mechanisms behind BAL formation and variability itself. In Section \ref{section:intro} we mentioned that the observed BAL-trough variability could be caused by variations in the covering factor which originate from the motion of the gaseous clouds along our line of sight; nevertheless, the cause of variability coordination in multiple BALs at different velocities in the same spectrum must likely arise from other mechanisms, as BALs arising at different velocities correspond to different radial distances from the central SMBH and therefore originate in gaseous regions that are separated from one another \citep[e.g.,][]{capellupo,nur12}. 

The cause of the variability must hence be global, rather than local; if we assume the existence of shielding gas between the radiation source and the wind, we can attribute coordinated variability to changes in the ionization level of the absorbing gas, originating from changes in the ionizing flux reaching the gas itself, which could be in turn ascribed to variations in the column density of the shielding gas. Such changes affect the outflow as a whole, thus giving rise to coordinated variations in the absorption troughs at different velocities. More saturated lines are scarcely responsive to changes in the ionization level, while changes in the covering factor can play a role in BAL variations; it is therefore likely that both causes contribute to the observed phenomenon, and the combined effect could be an enhanced variability or, in some cases, a partial balance \citep[e.g.,][]{capellupo, nur12}. Some recent works \citep[e.g., the aforementioned][]{baskin} tend to reject the shielding-gas scenario, rather favoring models where the changes in the ionization levels are an effect of radiation-pressure compression (see Section \ref{section:intro}). The work by \citet{saez} noted that changes in the shielding gas might cause significant variations in the X-ray emission from BAL QSOs, larger than the typical upper limits estimated for X-ray variability. They investigated the variability of 11 BAL QSOs over $3-30$ yr (rest frame), and such significant variations are not commonly observed. They thus infer that the shielding gas has rather stable properties on the timescales covered by their dataset. If we reject the shielding-gas hypothesis, we can likely ascribe BAL variability to changes in the ionization state of the extreme UV continuum, consistent with what is discussed in \citet{grier15}. However, there are still some proponents of the shielding-gas hypothesis, such as \citet{matthews}, and it is likely that the BAL variability phenomenon as a whole is the result of different causes.

We are planning to extend the analysis to lower ionization transitions --e.g., Si {\footnotesize{IV}} and Mg {\footnotesize{II}}-- in future works, as this would allow to study additional samples of BAL troughs and to investigate possible relations between the variability of troughs corresponding to different transitions. In addition, new spectra for our sample of BAL QSOs are currently being obtained by the SDSS-IV's Time Domain Spectroscopic Survey (TDSS; e.g., \citealt{tdss}), providing an opportunity for the analysis of re-emergence of previously disappeared BALs for those sources for which at least three epochs are available \citep[see, e.g.,][]{mcgraw}: these data would be a significant step towards a deeper understanding of the BAL phenomenon and could place additional --and possibly tighter-- constraints on the physics of BAL formation, evolution, and variability.

\begin{acknowledgements}
WNB and CJG thank NSF grants AST-1516784 and AST-1517113.\\
NFA thanks TUBITAK 115F037 for financial support.\\
Funding for SDSS-III has been provided by the Alfred P. Sloan Foundation, the Participating Institutions, the National Science Foundation, and the U.S. Department of Energy Office of Science. The SDSS-III Web site is http://www.sdss3.org/.\\
SDSS-III is managed by the Astrophysical Research Consortium for the Participating Institutions of the SDSS-III Collaboration including the University of Arizona, the Brazilian Participation Group, Brookhaven National Laboratory, University of Cambridge, Carnegie Mellon University, University of Florida, the French Participation Group, the German Participation Group, Harvard University, the Instituto de Astrofisica de Canarias, the Michigan State/Notre Dame/JINA Participation Group, Johns Hopkins University, Lawrence Berkeley National Laboratory, Max Planck Institute for Astrophysics, New Mexico State University, New York University, Ohio State University, Pennsylvania State University, University of Portsmouth, Princeton University, the Spanish Participation Group, University of Tokyo, University of Utah,Vanderbilt University, University of Virginia, University of Washington, and Yale University.
\end{acknowledgements}

\onecolumn
\begin{longtable}{c c c c c c c}\caption{QSOs in the $P_4$ sample. \\
Columns: (1): source identification number in this work; (2): source SDSS identification; (3): redshift from \citet{hw}; (4): \emph{i}-band magnitude from \citet{schn07}; (5) absolute \emph{i}-band magnitude from \citet{shen}; (6) plate, MJD, and fiber identifying each SDSS-I/II spectrum where a BAL trough is detected, and the corresponding BOSS spectrum (MJD $\geq 55176$) where disappearance occurs (in case of multiple spectra, we adopt the rule defined in Section \ref{section:disapp}); (7): number of BAL troughs detected in the same spectrum.}\label{tab:disapp_bals}\\
\hline \hline ID & SDSS ID & redshift & \emph{i} mag & $M_{i}$ & plate-MJD-fiber & $N_{tr.}$\\
  & & \emph{z} & (mag) & (mag) & &\\
(1) & (2) & (3) & (4) & (5) & (6) & (7)\\
\endfirsthead
\caption{Continued.} \\
\hline \hline ID & SDSS ID & redshift & \emph{i} mag & $M_{i}$ & plate-MJD-fiber & $N_{tr.}$\\
  & & \emph{z} & (mag) & (mag) & &\\
(1) & (2) & (3) & (4) & (5) & (6) & (7)\\
\hline
\endhead
\hline
\endfoot
\hline
\  28  & J004022.40+005939.6 & $2.5650\pm0.0006$ & $19.22\pm0.03$ & --26.751 & 0690-52261-563 & 1\\  
     &                                       &                              &                            &              & 3587-55182-950 & 0\\
\  94  & J021755.25-090141.0 & $2.3254\pm0.0004$ & $18.78\pm0.02$ & --26.975 & 0668-52162-163  & 2\\
     &                                       &                              &                            &              & 4395-55828-262 & 0\\
\  121 & J030004.75-063224.8 & $2.1821\pm0.0006$ & $18.98\pm0.02$ & --26.708 & 0458-51929-330 & 1\\
     &                                       &                              &                            &              & 7056-56577-060 & 0\\
\  235 & J081102.91+500724.2 & $1.8422\pm0.0006$ & $18.84\pm0.02$ & --26.445 & 0440-51912-395 & 1\\ 
     &                                       &                              &                            &              & 3699-55517-062 & 0\\
\  247 & J081338.34+240729.1 & $1.8273\pm0.0006$ & $18.43\pm0.01$ & --26.835 & 1585-52962-262 & 2\\
     &                                       &                              &                            &              & 4469-55863-924 & 1\\
\  407 & J090757.38+333116.2 & $1.9362\pm0.0005$ & $18.69\pm0.02$ & --26.643 & 1272-52989-543 & 2\\
     &                                       &                              &                            &              & 5812-56354-032 & 1\\
\  428  & J091159.36+442526.8 & $2.1713\pm0.0003$ & $17.79\pm0.02$ & --27.791 & 0832-52312-182 & 1\\
     &                                       &                              &                            &              & 4687-56338-654 & 0\\
\  447 & J091808.80+005457.7 & $2.116\pm0.001$ & $18.76\pm0.02$ & --26.783 & 0472-51955-615 & 1\\
     &                                       &                              &                            &              & 3821-55535-914 & 1\\
\  451 & J091944.53+560243.3 & $1.7964\pm0.0005$ & $18.75\pm0.02$ & --26.414 & 0451-51908-195 & 1\\
     &                                       &                              &                            &              & 5725-56625-675 & 0\\
\  469 & J092418.53+271851.5 & $2.1624\pm0.0005$ & $18.79\pm0.02$ & --26.806 & 1940-53383-352 & 2\\
     &                                       &                              &                            &              & 5797-56273-154 & 0\\
\  471 & J092444.66-000924.0 & $2.8632\pm0.0003$ & $19.22\pm0.02$ & --27.019 & 0474-52000-178 & 4\\
     &                                       &                              &                            &              & 3823-55534-262 & 1\\
\  490 & J092851.41+311627.0 & $2.0478\pm0.0005$ & $18.62\pm0.03$ & --26.835 & 1941-53386-168 & 3\\
     &                                       &                              &                            &              & 5807-56329-416 & 1\\
\  515 & J093418.28+355508.3 & $2.4402\pm0.0007$ & $18.90\pm0.01$ & --26.949 & 1275-52996-096 & 2\\
     &                                       &                              &                            &              & 4575-55590-498 & 1\\
\  526 & J093620.52+004649.2 & $1.7213\pm0.0005$ & $18.39\pm0.02$ & --26.746 & 0476-52314-444 & 1\\
     &                                       &                              &                            &              & 3826-55563-542 & 0\\
\  549 & J094437.56+104726.8 & $2.0334\pm0.0007$ & $18.89\pm0.02$ & --26.554 & 1742-53053-100 & 1\\
     &                                       &                              &                            &              & 5321-55945-498 & 0\\
\  565 & J094804.89+473223.0 & $1.7125\pm0.0005$ & $18.03\pm0.02$ & --26.998 & 1005-52703-121 & 2\\
     &                                       &                              &                            &              & 5741-55980-764 & 2\\
\  572 & J095035.10+560253.1 & $2.1775\pm0.0006$ & $19.06\pm0.02$ & --26.528 & 0557-52253-126 & 2\\
     &                                       &                              &                            &              & 5743-56011-644 & 1\\
\  603 & J100117.02+642009.4 & $1.9490\pm0.0006$ & $19.19\pm0.02$ & --26.179 & 0487-51943-565 & 1\\
     &                                       &                              &                            &              & 5722-56008-829 & 0\\
\  605 & J100131.95+053322.7 & $1.9594\pm0.0006$ & $18.76\pm0.02$ & --26.604 & 0995-52731-092 & 3\\
     &                                       &                              &                            &              & 4800-55674-678 & 0\\
\  623 & J100607.17+625320.2 & $1.8504\pm0.0005$ & $19.08\pm0.02$ & --26.143 & 0487-51943-075 & 3\\
     &                                       &                              &                            &              & 5722-56008-076 & 1\\
\  668 & J102250.16+483631.1 & $2.0601\pm0.0005$ & $18.64\pm0.02$ & --26.82 & 0873-52674-555 & 2\\
     &                                       &                              &                            &              & 7386-56769s584 & 1\\
\  673 & J102435.39+372637.0 & $1.7287\pm0.0005$ & $18.31\pm0.02$ & --26.741 & 1957-53415-515 & 1\\
     &                                       &                              &                            &              & 4559-55597-470 & 0\\
\  682 & J102812.08+381132.9 & $1.8196\pm0.0006$ & $19.08\pm0.02$ & --26.08 & 1428-52998-110 & 1\\
     &                                       &                              &                            &              & 4559-55597-654 & 0\\
\  704 & J103311.79+603146.5 & $2.5031\pm0.0003$ & $18.66\pm0.02$ & --27.217 & 0560-52296-402 & 1\\
     &                                       &                              &                            &              & 7090-56659-084 & 0\\
\  735 & J104509.67+480429.8 & $1.7834\pm0.0006$ & $19.07\pm0.02$ & --26.065 & 0963-52643-359 & 1\\
     &                                       &                              &                            &              & 6701-56367-432 & 0\\
\  752 & J104841.02+000042.8 & $2.0263\pm0.0006$ & $18.72\pm0.02$ & --26.759 & 0276-51909-310 & 2\\
     &                                       &                              &                            &              & 3835-55570-398 & 0\\
\  794 & J110038.71+450626.2 & $1.8711\pm0.0006$ & $19.04\pm0.03$ & --26.18 & 1436-53054-128 & 2\\
     &                                       &                              &                            &              & 4689-55656-910 & 1\\
\  808 & J110549.37+663456.8 & $2.0352\pm0.0006$ & $19.15\pm0.02$ & --26.296 & 0490-51929-557 & 2\\
     &                                       &                              &                            &              & 7111-56741-396 & 2\\
\  816 & J110906.31+640704.9 & $1.8555\pm0.0004$ & $17.31\pm0.02$ & --27.902 & 0596-52370-426 & 4\\
     &                                       &                              &                            &              & 7110-56746-430 & 0\\
\  885 & J112602.81+003418.2 & $1.7928\pm0.0005$ & $18.08\pm0.02$ & --27.073 & 0281-51614-432 & 2\\
     &                                       &                              &                            &              & 3839-55575-844 & 1\\
\  911  & J113236.06+030335.1 & $1.7609\pm0.0005$ & $18.47\pm0.01$ & --26.647 & 0513-51989-335 & 2\\
     &                                       &                              &                            &              & 4768-55944-124 & 1\\
\  917 & J113423.00+150059.2 & $2.1074\pm0.0004$ & $18.75\pm0.04$ & --26.791 & 1755-53386-396 & 2\\
     &                                       &                              &                            &              & 5373-56010-103 & 0\\
\  919 & J113438.58+091012.6 & $1.7967\pm0.0006$ & $18.65\pm0.02$ & --26.533 & 1224-52765-257 & 1\\
     &                                       &                              &                            &              & 5375-55973-304 & 0\\
\  928 & J113754.91+460227.4 & $2.1102\pm0.0006$ & $18.94\pm0.02$ & --26.606 & 1442-53050-004 & 1\\
     &                                       &                              &                            &              & 6647-56390-040 & 0\\
\  986 & J115244.20+030624.4 & $2.0865\pm0.0005$ & $18.31\pm0.02$ & --27.209 & 0515-52051-464 & 3\\
     &                                       &                              &                            &              & 4765-55674-228 & 1\\
\  1005 & J115707.36+333257.9 & $2.262\pm0.0003$ & $18.82\pm0.03$ & --26.859 & 2099-53469-197  & 1\\
     &                                       &                              &                            &              & 4647-55621-252 & 0\\
\  1184 & J124505.66+561430.5 & $2.1136\pm0.0004$ & $18.64\pm0.02$ & --26.855 & 1317-52765-202 & 1\\
     &                                       &                              &                            &              & 6832-56426-888 & 0\\
\  1203 & J125432.78+435228.9 & $2.1612\pm0.0006$ & $19.13\pm0.02$ & --26.441 & 1373-53063-052 & 1\\
     &                                       &                              &                            &              & 6619-56371-696 & 0\\
\  1235 & J130542.35+462503.4 & $1.8272\pm0.0005$ & $18.91\pm0.02$ & --26.267 & 1459-53117-165 & 2\\
     &                                       &                              &                            &              & 6624-56385-476 & 1\\
\  1252 & J131038.17+113617.9 & $1.6974\pm0.0006$ & $18.76\pm0.02$ & --26.269 & 1696-53116-027 & 1\\
     &                                       &                              &                            &              & 5422-55986-304 & 0\\
\  1269 & J131524.71+130411.8 & $2.3149\pm0.0005$ & $18.85\pm0.02$ & --26.893 & 1697-53142-525 & 1\\
     &                                       &                              &                            &              & 5425-56003-322 & 0\\
\  1286 & J132216.24+052446.3 & $2.0498\pm0.0006$ & $18.38\pm0.02$ & --27.09 & 0851-52376-622 & 4\\
     &                                       &                              &                            &              & 4761-55633-794 & 1\\
\  1298 & J132508.81+122314.2 & $1.7724\pm0.0005$ & $17.92\pm0.01$ & --27.217 & 1698-53146-509 & 1\\
     &                                       &                              &                            &              & 5432-56008-492 & 0\\
\  1316 & J133119.14+035658.0 & $1.7348\pm0.0005$ & $17.97\pm0.01$ & --27.127 & 0853-52374-295 & 2\\
     &                                       &                              &                            &              & 4759-55649-279 & 1\\
\  1323 & J133211.21+392825.9 & $2.0520\pm0.0009$ & $19.02\pm0.02$ & --26.402 & 2005-53472-330 & 1\\
     &                                       &                              &                            &              & 4708-55704-412 & 0\\
\  1359 & J134544.55+002810.7 & $2.4680\pm0.0005$ & $18.53\pm0.02$ & --27.349 & 0300-51943-382 & 1\\
     &                                       &                              &                            &              & 4043-55630-868 & 0\\
\  1393 & J135910.45+563617.3 & $2.2498\pm0.0006$ & $17.78\pm0.03$ & --27.868 & 1159-52669-296 & 2\\
     &                                       &                              &                            &              & 6801-56487-742 & 1\\
\  1400 & J140051.80+463529.9 & $1.9702\pm0.0005$ & $17.65\pm0.02$ & --27.695 & 1285-52723-104 & 1\\
     &                                       &                              &                            &              & 6750-56367-306 & 0\\
\  1408 & J140231.80+643610.4 & $1.9109\pm0.0005$ & $19.07\pm0.02$ & --26.192 & 0498-51984-238 & 2\\
     &                                       &                              &                            &              & 6986-56717-170 & 0\\
\  1414 & J140501.94+444759.7 & $2.2149\pm0.0004$ & $17.86\pm0.04$ & --27.751 & 1467-53115-494 & 3\\
     &                                       &                              &                            &              & 6055-56102-576 & 1\\
\  1441 & J141407.25+562010.3 & $2.2822\pm0.0003$ & $18.98\pm0.02$ & --26.706 & 1160-52674-231 & 1\\
     &                                       &                              &                            &              & 6803-56402-426 & 0\\
\  1464 & J142132.01+375230.3 & $1.7791\pm0.0006$ & $18.66\pm0.02$ & --26.73 & 1380-53084-013 & 1\\
     &					    &				   &.                           &              & 4713-56044-532 & 0\\
\  1467 & J142140.27-020239.0 & $2.0878\pm0.0006$ & $18.88\pm0.02$ & --26.68 & 0917-52400-546 & 3\\
     &                                       &                              &                            &              & 4032-55333-736 & 1\\
\  1480 & J142514.60+632703.8 & $2.1767\pm0.0003$ & $18.13\pm0.01$ & --27.465 & 0499-51988-179 & 4\\
     &                                       &                              &                            &              & 7124-56720-431 & 0\\
\  1493 & J142813.72+233742.8 & $1.9383\pm0.0006$ & $18.64\pm0.02$ & --26.752 & 2136-53494-360 & 1\\
     &                                       &                              &                            &              & 6014-56072-778 & 0\\
\  1534 & J143821.60+393407.3 & $2.0430\pm0.0007$ & $18.66\pm0.01$ & --26.78 & 1349-52797-022 & 2\\
     &                                       &                              &                            &              & 5171-56038-070 & 1\\
\  1638 & J151610.07+434506.7 & $1.8218\pm0.0005$ & $18.47\pm0.01$ & --26.708 & 1677-53148-599 & 1\\
     &                                       &                              &                            &              & 6040-56101-448 & 0\\
\  1650 & J152149.78+010236.4 & $2.2386\pm0.0004$ & $18.56\pm0.02$ & --27.159 & 0313-51673-339 & 1\\
     &                                       &                              &                            &              & 4011-55635-166 & 0\\
\  1651 & J152243.98+032719.8 & $2.0002\pm0.0005$ & $18.65\pm0.02$ & --26.803 & 0592-52025-254 & 1\\
     &                                       &                              &                            &              & 4803-55734-442 & 0\\
\  1693 & J154256.06+372746.4 & $1.7440\pm0.0006$ & $19.11\pm0.02$ & --25.97 & 1416-52875-380 & 2\\
     &                                       &                              &                            &              & 4973-56042-533 & 1\\
\  1701 & J154621.25+521303.4 & $2.7794\pm0.0003$ & $19.07\pm0.02$ & --27.057 & 0618-52049-271 & 3\\
     &                                       &                              &                            &              & 6715-56449-234 & 2\\
\  1702 & J154655.55+370739.2 & $1.8022\pm0.0006$ & $18.93\pm0.02$ & --26.212 & 1416-52875-529 & 1\\
     &                                       &                              &                            &              & 4973-56042-318 & 0\\
\  1715 & J155119.14+304019.8 & $2.4104\pm0.0004$ & $18.49\pm0.02$ & --27.362 & 1580-53145-008 & 2\\
     &                                       &                              &                            &              & 5011-55739-054 & 1\\
\  1717 & J155135.78+464609.4 & $1.8975\pm0.0007$ & $18.85\pm0.02$ & --26.389 & 1168-52731-115 & 2\\
     &                                       &                              &                            &              & 6730-56425-366 & 1\\
\  1823 & J163844.42+350857.4 & $2.2635\pm0.0007$ & $18.92\pm0.02$ & --26.784 & 1339-52767-602 & 2\\
     &                                       &                              &                            &              & 5188-55803-876 & 2\\
\  1824 & J163847.42+232716.4 & $3.864\pm0.002$ & $18.40\pm0.02$ & --28.506 & 1571-53174-539 & 2\\
     &                                       &                              &                            &              & 4186-55691-408 & 1\\
\  1853 & J165543.23+394519.8 & $1.7545\pm0.0005$ & $17.90\pm0.01$ & --27.197 & 0633-52079-353 & 2\\
     &                                       &                              &                            &              & 6063-56098-438 & 1\\
\hline
\end{longtable}

\newpage
\begin{longtable}{c c c c r r r c c}
\caption{Parameters of disappearing BAL troughs. \\
Columns: (1): source identification number in this work; (2): source SDSS identification; (3): MJD of the SDSS-I/II epoch where the EW of the disappearing BAL trough is measured (i.e., the latest SDSS-I/II epoch where the BAL trough is detected); (4): rest-frame EW measurement; (5): maximum velocity of the BAL trough; (6) minimum velocity of the BAL trough; (7) rest-frame $\Delta t$ between the SDSS-I/II epoch where the BAL trough is detected and the BOSS epoch where disappearance of the same BAL is observed; (8) 1 = BALs belonging to the $P_8$ sample, 0 = BALs not belonging to the $P_8$ sample; (9) 1 = BALs belonging to the pristine sample, 0 = BALs not belonging to the pristine sample.}\label{tab:disapp_bals1}\\
\hline \hline ID & SDSS ID & MJD & EW & $v_{max}$ & $v_{min}$ & Rest-frame $\Delta t$ & $P_8$ sample & Pristine sample\\
  & & & (\AA) & (km s$^{-1}$) & (km s$^{-1}$) & (days) & & \\
(1) & (2) & (3) & (4) & (5) & (6) & (7) & (8) & (9)\\
\endfirsthead
\caption{Continued.} \\
\hline \hline ID & SDSS ID & MJD & EW & $v_{max}$ & $v_{min}$ & Rest-frame $\Delta t$ & $P_8$ sample & Pristine sample\\
  & & & (\AA) & (km s$^{-1}$) & (km s$^{-1}$) & (days) & & \\
(1) & (2) & (3) & (4) & (5) & (6) & (7) & (8) & (9)\\
\hline
\endhead
\hline
\endfoot
\hline
\  28 & J004022.40+005939.6 & 52261 & $10.67\pm0.59$ & --9514 & --3719 & 819.35 & 1 & 1\\
\  94 & J021755.25-090141.0 & 52162 & $6.43\pm0.31$ & --22099 & --16393 & 1102.42 & 1 & 0\\
\ 121 & J030004.75-063224.8 & 51929 & $4.15\pm0.38$ & --23999 & --20979 & 1460.67 & 0 & 1\\
\ 235 & J081102.91+500724.2 & 51912 & $5.77\pm0.47$ & --11855 & --8959 & 1268.38 & 1 & 1\\
\ 247 & J081338.34+240729.1 & 52962 & $2.59\pm0.23$ & --9729 & --6970 & 1026.07 & 1 & 1\\
\ 407 & J090757.38+333116.2 & 52989 & $2.65\pm0.21$ & --24803 & --22608 & 1146.04 & 0 & 1\\
\ 428 & J091159.36+442526.8 & 52312 & $4.82\pm0.15$ & --21064 & --15563 & 1269.51 & 1 & 0\\
\ 447 & J091808.80+005457.7 & 51955 & $11.54\pm0.65$ & --26935 & --22067 & 1148.91 & 1 & 1\\
\ 451 & J091944.53+560243.3 & 51908 & $3.85\pm0.24$ & --14090 & --11403 & 1686.81 & 1 & 1\\
\ 469 & J092418.53+271851.5 & 53383 & $4.46\pm0.15$ & --8933 & --6381 & 913.86 & 1 & 1\\
\ 471 & J092444.66-000924.0 & 52000 & $4.55\pm0.47$ & --27243 & --24708 & 914.79 & 0 & 1\\
\ 490 & J092851.41+311627.0 & 53386 & $4.60\pm0.22$ & --23706 & --20342 & 965.61 & 1 & 0\\
\ 515 & J093418.28+355508.3 & 52996 & $2.97\pm0.17$ & --25007 & --22332 & 754.03 & 1 & 0\\
\ 526 & J093620.52+004649.2 & 52314 & $6.36\pm0.34$ & --17103 & --13179 & 1193.91 & 1 & 1\\
\ 549 & J094437.56+104726.8 & 53053 & $12.43\pm0.45$ & --26135 & --19547 & 953.39 & 1 & 0\\
\ 565 & J094804.89+473223.0 & 52703 & $4.25\pm0.26$ & --16548 & --14139 & 1208.11 & 1 & 0\\
\ 572 & J095035.10+560253.1 & 52253 & $3.10\pm0.31$ & --24735 & --22608 & 1182.69 & 0 & 0\\
\ 603 & J100117.02+642009.4 & 51943 & $9.69\pm0.49$ & --15313 & --9111 & 1378.43 & 1 & 1\\
\ 605 & J100131.95+053322.7 & 52731 & $9.27\pm0.42$ & --14851 & --6992 & 994.46 & 1 & 0\\
\ 623 & J100607.17+625320.2 & 51943 & $3.84\pm0.24$ & --8167 & --6097 & 1426.12 & 1 & 0\\
\ 668 & J102250.16+483631.1 & 52674 & $2.63\pm0.18$ & --15424 & --12806 & 1338.19 & 0 & 0\\
\ 673 & J102435.39+372637.0 & 53415 & $4.90\pm0.22$ & --10339 & --6200 & 799.65 & 1 & 1\\
\ 682 & J102812.08+381132.9 & 52998 & $3.95\pm0.31$ & --16423 & --13600 & 921.76 & 0 & 0\\
\ 704 & J103311.79+603146.5 & 52296 & $6.36\pm0.22$ & --23340 & --18669 & 1245.47 & 1 & 0\\
\ 735 & J104509.67+480429.8 & 52643 & $6.89\pm0.39$ & --3661 & --623 & 1337.93 & 1 & 0\\
\ 752 & J104841.02+000042.8 & 51909 & $2.29\pm0.14$ & --12239 & --10171 & 1209.73 & 1 & 1\\
\ 794 & J110038.71+450626.2 & 53054 & $14.09\pm0.53$ & --22239 & --13849 & 906.27 & 1 & 0\\
\ 808 & J110549.37+663456.8 & 51929 & $2.60\pm0.20$ & --26928 & --24256 & 1585.40 & 1 & 1\\
\ 816 & J110906.31+640704.9 & 52370 & $3.96\pm0.16$ & --24046 & --19308 & 1532.48 & 1 & 1\\
\ 816 & J110906.31+640704.9 & 52370 & $2.68\pm0.11$ & --15593 & --12907 & 1532.48 & 1 & 1\\               
\ 885 & J112602.81+003418.2 & 51614 & $3.83\pm0.15$ & --25868 & --22439 & 1418.29 & 1 & 0\\
\ 911 & J113236.06+030335.1 & 51989 & $3.64\pm0.20$ & --13506 & --10404 & 1432.50 & 1 & 0\\
\ 917 & J113423.00+150059.2 & 53386 & $3.49\pm0.22$ & --7609 & --5125 & 844.44 & 1 & 0\\
\ 919 & J113438.58+091012.6 & 52765 & $3.83\pm0.32$ & --6746 & --4675 & 1147.07 & 1 & 0\\
\ 928 & J113754.91+460227.4 & 53050 & $7.21\pm0.37$ & --14358 & --9948 & 1073.89 & 1 & 1\\
\ 986 & J115244.20+030624.4 & 52051 & $7.54\pm0.22$ & --15307 & --10070 & 1173.82 & 1 & 1\\
\ 1005 & J115707.36+333257.9 & 53469 & $3.46\pm0.24$ & --14218 & --11324 & 659.72 & 0 & 0\\
\ 1184 & J124505.66+561430.5 & 52765 & $4.05\pm0.26$ & --17713 & --13996 & 1175.81 & 0 & 0\\
\ 1203 & J125432.78+435228.9 & 53063 & $4.47\pm0.28$ & --9371 & --6681 & 1046.44 & 1 & 1\\
\ 1235 & J130542.35+462503.4 & 53117 & $5.25\pm0.31$ & --12683 & --8960 & 1155.91 & 1 & 1\\
\ 1235 & J130542.35+462503.4 & 53117 & $3.55\pm0.24$ & --8753 & --6546 & 1155.91 & 1 & 1\\
\ 1252 & J131038.17+113617.9 & 53116 & $6.54\pm0.48$ & --18252 & --14397 & 1063.99 & 0 & 1\\
\ 1269 & J131524.71+130411.8 & 53142 & $8.04\pm0.21$ & --10211 & --5865 & 863.07 & 1 & 0\\
\ 1286 & J132216.24+052446.3 & 52376 & $2.63\pm0.16$ & --22186 & --19919 & 1067.94 & 1 & 0\\
\ 1286 & J132216.24+052446.3 & 52376 & $2.84\pm0.17$ & --18132 & --15311 & 1067.94 & 0 & 0\\
\ 1286 & J132216.24+052446.3  & 52376 & $2.25\pm0.15$ & --12831 & --10763 & 1067.94 & 1 & 0\\
\ 1298 & J132508.81+122314.2 & 53146 & $13.08\pm0.24$ & --9787 & --4820 & 1032.32 & 1 & 0\\
\ 1316 & J133119.14+035658.0 & 52374 & $3.17\pm0.33$ & --24008 & --21399 & 1197.53 & 0 & 0\\
\ 1323 & J133211.21+392825.9 & 53472 & $4.78\pm0.27$ & --20822 & --17384 & 731.32 & 1 & 0\\
\ 1359 & J134544.55+002810.7 & 51943 & $16.46\pm0.27$ & --12285 & --5457 & 1063.15 & 1 & 0\\
\ 1393 & J135910.45+563617.3 & 52669 & $7.26\pm0.21$  & --28071 & --19772 & 1174.84 & 1 & 0\\ 
\ 1400 & J140051.80+463529.9 & 52723 & $6.36\pm0.16$ & --19586 & --14770 & 1226.85 & 1 & 0\\
\ 1408 &  J140231.80+643610.4 & 51984 & $6.51\pm0.42$ & --20508 & --17070 & 1625.96 & 1 & 0\\
\ 1414 & J140501.94+444759.7 & 53115 & $2.44\pm0.12$ & --23007 & --19986 & 929.11 & 1 & 1\\
\ 1441 & J141407.25+562010.3 & 52674 & $6.05\pm0.23$ & --19228 & --15513 & 1135.82 & 1 & 1\\ 
\ 1464 & J142132.01+375230.3 & 53084 & $5.27\pm0.34$ & --16435 & --13680 & 1065.09 & 0 & 0\\
\ 1467 & J142140.27-020239.0 & 52400 & $4.67\pm0.17$ & --15433 & --12471 & 949.87 & 1 & 0\\
\ 1480 & J142514.60+632703.8 & 51988 & $3.63\pm0.13$ & --23013 & --18204 & 1489.60 & 0 & 0\\ 
\ 1480 & J142514.60+632703.8 & 51988 & $2.42\pm0.10$ & --16140 & --13110 & 1489.60 & 0 & 0\\ 
\ 1493 & J142813.72+233742.8 & 53494 & $4.21\pm0.27$ & --11952 & --9125 & 877.38 & 1 & 0\\
\ 1534 & J143821.60+393407.3 & 52797 & $2.71\pm0.24$ & --27007 & --24951 & 1065.07 & 0 & 0\\
\ 1638 & J151610.07+434506.7 & 53148 & $4.98\pm0.15$ & --8663 & --6456 & 1046.50 & 1 & 0\\
\ 1650 & J152149.78+010236.4 & 51673 & $4.57\pm0.28$ & --22657 & --18741 & 1223.37 & 1 & 1\\
\ 1651 & J152243.98+032719.8 & 52025 & $2.83\pm0.20$ & --14056 & --11852 & 1236.25 & 1 & 1\\
\ 1693 & J154256.06+372746.4 & 52875 & $6.75\pm0.49$ & --16627 & --12840 & 1154.15 & 1 & 0\\
\ 1701 & J154621.25+521303.4 & 52049 & $3.93\pm0.47$ & --20573 & --18374 & 1164.21 & 0 & 0\\
\ 1702 & J154655.55+370739.2 & 52875 & $4.50\pm0.33$ & --12574 & --9817 & 1130.18 & 1 & 1\\
\ 1715 & J155119.14+304019.8 & 53145 & $7.01\pm0.30$ & --22347 & --17605 & 760.61 & 1 & 0\\
\ 1717 & J155135.78+464609.4 & 52731 & $2.37\pm0.27$ & --14658 & --12522 & 1274.89 & 0 & 0\\
\ 1823 & J163844.42+350857.4 & 52767 & $5.83\pm0.27$ & --21165 & --16971 & 930.29 & 1 & 1\\
\ 1824 & J163847.42+232716.4 & 53174 & $3.77\pm0.15$ & --19753 & --16796 & 517.43 & 1 & 1\\  
\ 1853 & J165543.23+394519.8 & 52079 & $8.13\pm0.26$ & --21549 & --16668 & 1459.07 & 1 & 1\\  
\hline
\end{longtable}
\twocolumn

\bibliographystyle{aa}
\bibliography{AA_2017_32497}{}

\end{document}